\documentclass{appolb}
\usepackage{graphicx}

\begin{document}
\title{
Double parton scattering at high energies
\thanks{Presented at the XXI Cracow EPIPHANY Conference on 
Future High Energy Colliders}%
}
\author
{Antoni Szczurek
\address{Institute of Nuclear Physics PAN, PL-31-342 Cracow, Poland and\\
University of Rzesz\'ow, PL-35-959 Rzesz\'ow, Poland}
}
\maketitle

\begin{abstract}
We discuss a few examples of rich newly developing field of double
parton scattering. We start our presentation from production
of two pairs of charm quark-antiquark and argue that it is the golden
reaction to study the double parton scattering effects. 
In addition to the DPS we consider briefly also mechanism of single parton
scattering and show that it gives much smaller contribution
to the $c \bar c c \bar c$ final state.
Next we discuss a perturbative parton-splitting mechanism which 
should be included in addition to the conventional DPS mechanism. 
We show that the presence
of this mechanism unavoidably leads to collision energy and other
kinematical variables dependence of so-called $\sigma_{eff}$ parameter 
being extracted from different experiments. Next we briefly discuss
production of four jets. We concentrate on estimation of the
contribution of DPS for jets remote in rapidity. Understanding of 
this contribution is very
important in the context of searches for BFKL effects known under 
the the name Mueller-Navelet jets.
We discuss the situation in a more general context.
Finally we briefly mention about DPS effects in production of $W^+ W^-$.
Outlook closes the presentation.
\end{abstract}


\section{Introduction}

It is well known that the multi-parton interaction in general
and double parton scattering processes in particular become more 
and more important at high energies. In the present short review 
we concentrate on double parton processes (DPS) which can be described 
as perturbative processes, i.e. processes where the hard scale 
is well defined
(production of heavy objects, or objects with large transverse
momenta). In general, the cross section for the double-parton scattering
grows faster than the corresponding (the same final state) cross section for single parton
scattering (SPS).

The double-parton scattering was recognized already 
in seventies and eighties 
\cite{LP78,T1979,GSH80,H1983,PT1984,PT1985,M1985,HO1985,SZ1987}. 
The activity stopped when it was realized that their contribution at 
the center-of-mass energies available at those times was negligible.
Several estimates of the cross section for different processes have been
presented in recent years \cite{DH1996,KS2000,FT2002,BJS2009,GKKS2010,
SV2011,BDFS2011,KKS2011,BSZ2011}. The theory of the double-parton 
scattering is quickly developing
(see e.g. \cite{S2003,KS2004,SS2004,GS2010,GS2011,DS2011,RS2011,DOS2011}).

In Ref.~\cite{LMS2012} we showed that the production of $c \bar c c \bar c$
is a very good place to study DPS effects.
Here, the quark mass is small enough to assure that the cross section 
for DPS is very large,
and large enough that each of the scatterings can be treated
within pQCD. 

The calculation performed in Ref.~\cite{LMS2012} were done in 
the leading-order (LO) collinear approximation. This may not be 
sufficient when comparing the results 
of the calculation with real experimental data. In the meantime 
the LHCb collaboration presented new interesting data for simultaneous 
production of two charmed mesons \cite{Aaij:2012dz}. 
They have observed large percentage of the events with two mesons, 
both containing $c$ quark, with respect to the typical production 
of the corresponding meson/antimeson pair 
($\sigma_{D_{i}D_{j}}/ \sigma_{D_{i}\bar{D_{j}}} \sim 10\%$).

In Ref.~\cite{MS2013} we discussed that the large effect is a 
footprint of double parton scattering. In this paper each step
of the double parton scattering was considered in the
$k_t$-factorization approach.
In Ref.~\cite{Berezhnoy2012} the authors estimated DPS contribution
based on the experimental inclusive $D$ meson spectra measured at the LHC.
In their approach fragmentation was included only in terms
of the branching fractions for the quark-to-meson transition $c \to D$. In our approach
in Ref.~\cite{MS2013} we included full kinematics of hadronization process.
There we showed also first differential distributions
on the hadron level that can be confronted with recent LHCb experimental
data \cite{Aaij:2012dz}.

\begin{figure}
\begin{center}
\includegraphics[width=5cm]{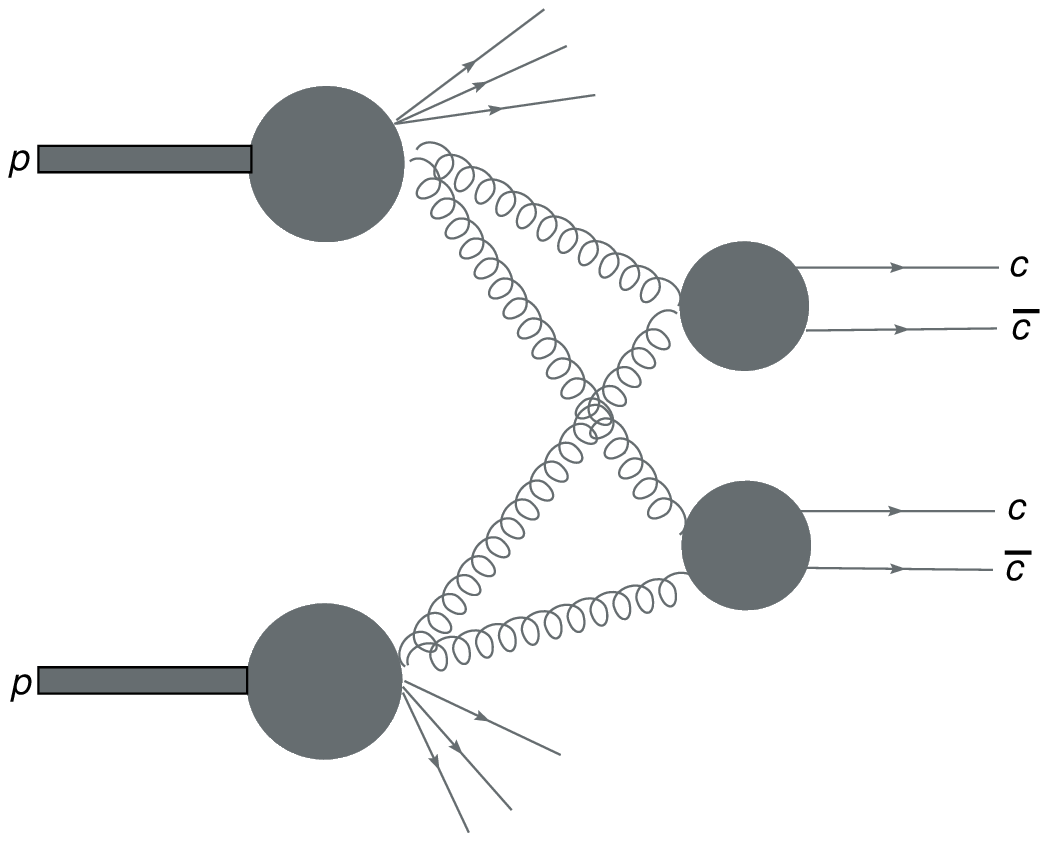}
\includegraphics[width=4cm]{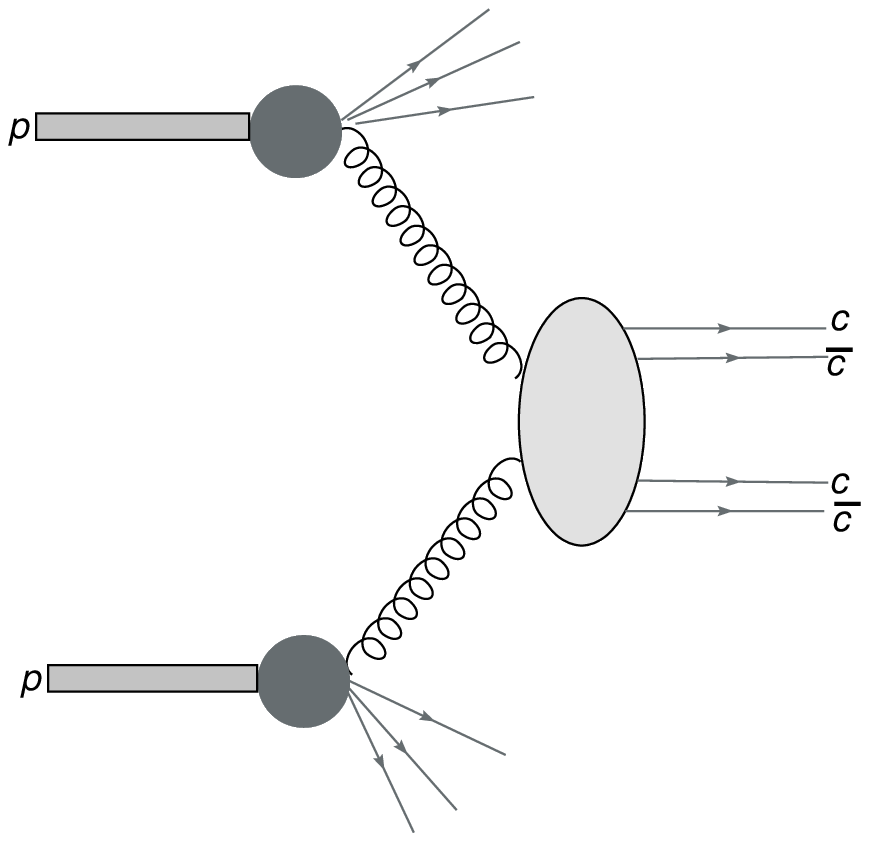}
\end{center}
\caption{
SPS and DPS production mechanisms of $c \bar c c \bar c$. 
}
\label{fig:diagrams:ccbarccbar}
\end{figure}

\begin{figure}[!h]
\begin{center}
\includegraphics[width=4.0cm]{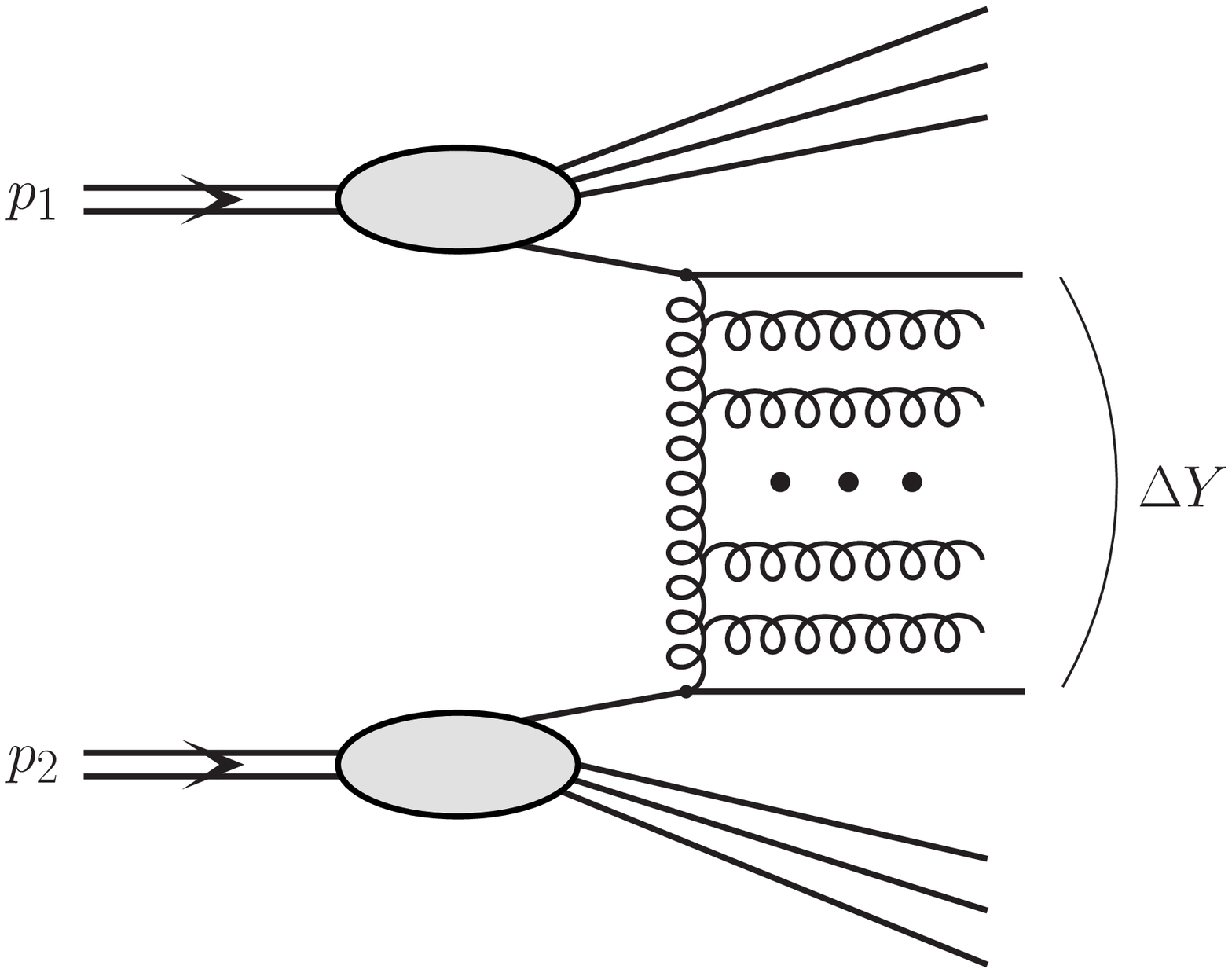}
\includegraphics[width=4.0cm]{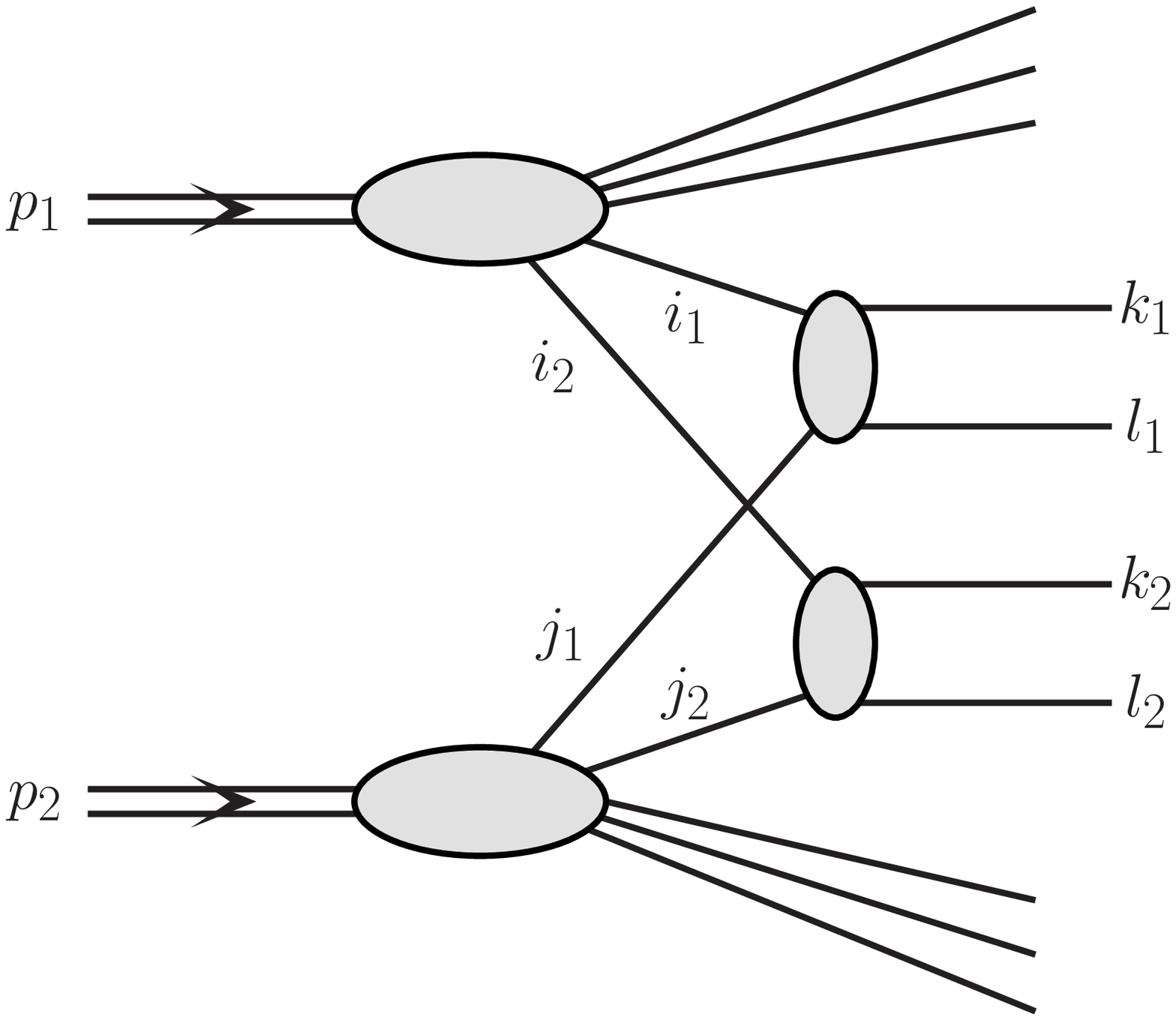}
\end{center}
\caption{
\small A diagramatic representation of the Mueller-Navelet jet
production (left diagram) and of the double paron scattering mechanism
(right diagram).
}
 \label{fig:diagrams}
\end{figure}

25 years ago Mueller and Navelet predicted strong decorrelation 
in relative azimuthal angle \cite{Mueller:1986ey} of jets with large
rapidity separation due to exchange of the BFKL ladder between quarks. 
The generic picture is presented in diagram (a) of Fig.~\ref{fig:diagrams}.
In a bit simplified picture quark/antiquarks are emitted forward
and backward, whereas gluons emitted along the ladder populate 
rapidity regions in between.
Due to diffusion along the ladder the correlation
between the most forward and the most backward jets is small.
This was a picture obtained within leading-logarithmic 
BFKL formalism
\cite{Mueller:1986ey,DelDuca:1993mn,Stirling:1994he,DelDuca:1994ng,Kim96,Andersen2001}.
Calculations of higher-order BFKL effects slightly modified
this simple picture 
\cite{Bartels-MNjets,Vera:2007kn,Marquet:2007xx,Colferai:2010wu,Caporale:2011cc,Ivanov:2012ms,
Caporale:2012ih,Ducloue:2013hia,Ducloue:2013bva,DelDuca2014} 
leading to smaller decorrelation
in rapidity. Recently the NLL corrections were calculated
both to the Green's function and to the jet vertices.
The effect of the NLL correction is large and leads to significant
lowering of the cross section.
So far only averaged values of $<\!cos(n \phi_{jj}\!>$ 
over available phase space or even their ratios 
were studied experimentally \cite{CMS_MN1}. More detailed studies 
are necessary to verify this type of calculations. In particular, 
the approach should reproduce dependence on the rapidity distance between
the jets emitted in opposite hemispheres.
Large-rapidity-distance jets can be produced only at high energies
where the rapidity span is large.
A first experimental search for the Mueller-Navelet jets was made by 
the D0 collaboration. In their study rapidity distance between jets was
limited to 5.5 units only.
In spite of this they have observed a broadening of the $\phi_{jj}$
distribution with growing rapidity distance between jets.
The dijet azimuthal correlations were also studied in collinear
next-to-leading order approximation \cite{Aurenche:2008dn}.
The LHC opens new possibility to study the decorrelation effect. 
First experimental data measured at $\sqrt{s}$ = 7 TeV are expected 
soon \cite{CMS_private}.

The double parton scattering mechanism of $W^+ W^-$ production
was discussed e.g. in 
Refs.~\cite{KS2000,Kulesza2010,GKKS2011,LSR2015,GL2014}. 
The $W^+ W^-$ final states constitutes a background to Higgs production.
It was discussed recently that double-parton scattering could explain
a large part of the observed signal \cite{KP2013}. We shall also discuss
the double parton scattering mechanism of $W^+W^-$ production in the present paper.

\section{Formalism used in the calculations}

\subsection{$c \bar c c \bar c$ production}

Let us consider first production of $c \bar c c \bar c$ final state
within the DPS framework. In a simple probabilistic picture the cross section for double-parton 
scattering can be written as:
\begin{equation}
\sigma^{DPS}(p p \to c \bar c c \bar c X) = \frac{1}{2 \sigma_{eff}}
\sigma^{SPS}(p p \to c \bar c X_1) \cdot \sigma^{SPS}(p p \to c \bar c X_2).
\label{basic_formula}
\end{equation}
This formula assumes that the two subprocesses are not correlated.
At low energies one has to include parton momentum conservation
i.e. extra limitations: $x_1+x_3 <$ 1 and $x_2+x_4 <$ 1, where 
$x_1$ and $x_3$ are longitudinal momentum fractions of gluons emitted 
from one proton and $x_2$ and $x_4$ their counterpairs for gluons 
emitted from the second proton. 
Experimental data \cite{Tevatron} provide an estimate of $\sigma_{eff}$
in the denominator of formula (\ref{basic_formula}). In our studies presented here we
usually take $\sigma_{eff}$ = 15 mb.


The simple formula (\ref{basic_formula}) can be generalized to address 
differential distributions. In leading-order approximation 
differential distribution can be written as
\begin{equation}
\frac{d \sigma}{d y_1 d y_2 d^2 p_{1t} d y_3 d y_4 d^2 p_{2t}}  \\ =
\frac{1}{ 2 \sigma_{eff} }
\frac{ d \sigma } {d y_1 d y_2 d^2 p_{1t}} \cdot
\frac{ d \sigma } {d y_3 d y_4 d^2 p_{2t}} 
\label{differential_distribution}
\end{equation}
which by construction reproduces formula for integrated cross section 
(\ref{basic_formula}).
This cross section is formally differential in 8 dimensions but can be 
easily reduced to 7 dimensions noting that physics of unpolarized
scattering cannot depend on azimuthal angle of the pair or on azimuthal 
angle of one of the produced $c$ ($\bar c$) quark (antiquark).
The differential distributions for each single scattering step can 
be written in terms of collinear gluon distributions with 
longitudinal momentum fractions
$x_1$, $x_2$, $x_3$ and $x_4$ expressed in terms of rapidities 
$y_1$, $y_2$, $y_3$, $y_4$ and transverse momenta of quark (or antiquark)
for each step 
(in the LO approximation identical for quark and antiquark).

A more general formula for the cross section can be written formally 
in terms of double-parton distributions, e.g. $F_{gg}$, $F_{qq}$, etc. 
In the case of heavy quark (antiquark) production at high energies:
\begin{eqnarray}
d \sigma^{DPS} &=& \frac{1}{2 \sigma_{eff}}
F_{gg}(x_1,x_2,\mu_1^2,\mu_2^2) F_{gg}(x'_{1}x'_{2},\mu_1^2,\mu_2^2)
\nonumber \\
&&d \sigma_{gg \to c \bar c}(x_1,x'_{1},\mu_1^2)
d \sigma_{gg \to c \bar c}(x_2,x'_{2},\mu_2^2) \; dx_1 dx_2 dx'_1 dx'_2 \, .
\label{cs_via_doublePDFs}
\end{eqnarray}
It is rather inspiring to write the double-parton distributions 
in the impact parameter space $F_{gg}(x_1,x_2,b) = g(x_1) g(x_2) F(b)$, 
where $g$ are usual conventional parton distributions and $F(b)$ is 
an overlap of the matter distribution in the 
transverse plane where $b$ is a distance between both gluons in the
transverse plane \cite{CT1999}. The effective cross section in 
(\ref{basic_formula}) is then $1/\sigma_{eff} = \int d^2b F^2(b)$ and 
in this approximation is energy independent.

Even if the factorization is valid at some scale, QCD evolution may lead
to a factorization breaking. Evolution is known only in the case 
when the scale of both scatterings is the same \cite{S2003,KS2004,GS2010}
i.e. for heavy object, like double gauge boson production.

In Ref.~\cite{LMS2012} we applied the commonly used in the literature
factorized model to $p p \to c \bar c c \bar c$ and predicted
that at the LHC energies the cross section for two $c \bar c$ pair
production starts to be of the same size as that for single $c \bar c$
production.

In LO collinear approximation the differential distributions for $c\bar{c}$ production
depend e.g. on rapidity of quark, rapidity of antiquark and transverse
momentum of one of them \cite{LMS2012}. 
In the next-to-leading order (NLO) collinear approach or in the
$k_t$-factorization approach the situation is more complicated as there 
are more kinematical variables needed to describe the kinematical situation.
In the $k_t$-factorization approach the differential cross section for
DPS production of $c \bar c c \bar c$ system, assuming factorization of 
the DPS model, can be written as: 
\begin{eqnarray}
\frac{d \sigma^{DPS}(p p \to c \bar c c \bar c X)}{d y_1 d y_2 d^2 p_{1,t} d^2 p_{2,t} 
d y_3 d y_4 d^2 p_{3,t} d^2 p_{4,t}} = \nonumber \;\;\;\;\;\;\;\;\;\;\;\;\;\;\;\;\;\;\;\;\;\;\;\;\;\;\;\;\;\;\;\;\;\;\;\;\;\;\;\;\;\;\;\;\\ 
\frac{1}{2 \sigma_{eff}} \cdot
\frac{d \sigma^{SPS}(p p \to c \bar c X_1)}{d y_1 d y_2 d^2 p_{1,t} d^2 p_{2,t}}
\cdot
\frac{d \sigma^{SPS}(p p \to c \bar c X_2)}{d y_3 d y_4 d^2 p_{3,t} d^2 p_{4,t}}.
\end{eqnarray}

Again when integrating over kinematical variables one recovers Eq.(\ref{basic_formula}).
%
%
%
\begin{equation}
\sigma_{eff} = \left[ \int d^{2}b \; (T(\vec{b}))^{2} \right]^{-1},
\end{equation}
where the overlap function
\begin{equation}
T ( \vec{b} ) = \int f( \vec{b}_{1} ) f(\vec{b}_{1} - \vec{b} ) \; d^2 b_{1},
\end{equation}
if the impact-parameter dependent double-parton distributions (dPDFs) 
are written in the following factorized approximation 
\cite{GS2010,Gustaffson2011}:
\begin{equation}
\Gamma_{i,j}(x_1,x_2;\vec{b}_{1},\vec{b}_{2};\mu_{1}^{2},\mu_{2}^{2}) = F_{i,j}(x_1,x_2;\mu_{1}^{2},\mu_{2}^{2}) \; f(\vec{b}_{1}) \; f(\vec{b}_{2}).
\end{equation}
Without loosing generality the impact-parameter distribution
can be written as
\begin{equation}
\Gamma(b,x_1,x_2;\mu_1^2,\mu_2^2) = F(x_1,\mu_1^2) \; F(x_2,\mu_2^2) \;
F(b;x_1,x_2,\mu_1^2,\mu_2^2), \;
\label{correlation_function}
\end{equation}
where $b$ is the parton separation in the impact parameters space.
In the formula above the function $F(b;x_1,x_2,\mu_1^2,\mu_2^2)$
contains all information about correlations between the two partons
(two gluons in our case). The dependence was studied
numerically in Ref.~\cite{Gustaffson2011} within Lund Dipole Cascade model.
The biggest discrepancy was found in the small $b$ region,
particularly for large $\mu_1^2$ and/or $\mu_2^2$. We shall return
to the issue when commenting our results.
In general the effective cross section may
depend on many kinematical variables:
\begin{equation}
\sigma_{eff}(x_1,x_2,x'_1,x'_2,\mu_1^2,\mu_2^2) =
\left(
\int d^2 b \; F(b;x_1,x_2,\mu_1^2,\mu_2^2) \; F(b;x'_1,x'_2,\mu_1^2,\mu_2^2)
\right)^{-1}.
\label{generalized_sigma_eff}
\end{equation}
We shall return to these dependences when discussing the role of
perturbative parton splitting.

\begin{figure}
\begin{center}
\includegraphics[width=7cm]{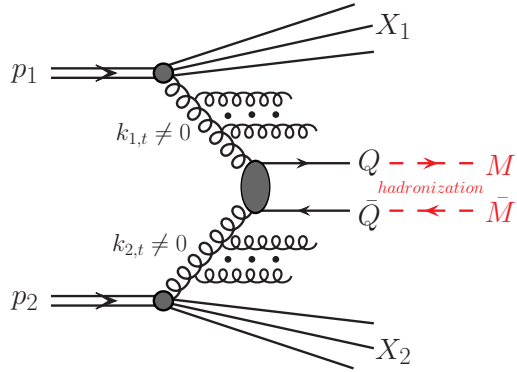}
\end{center}
\caption{
Production of $c \bar c$ quark and antiquark
via fusion of virtual gluons.
}
\label{gg_ccbar}
\end{figure}

\subsection{Parton splitting}

\begin{figure}[!h]
\begin{center}
\includegraphics[width=4cm]{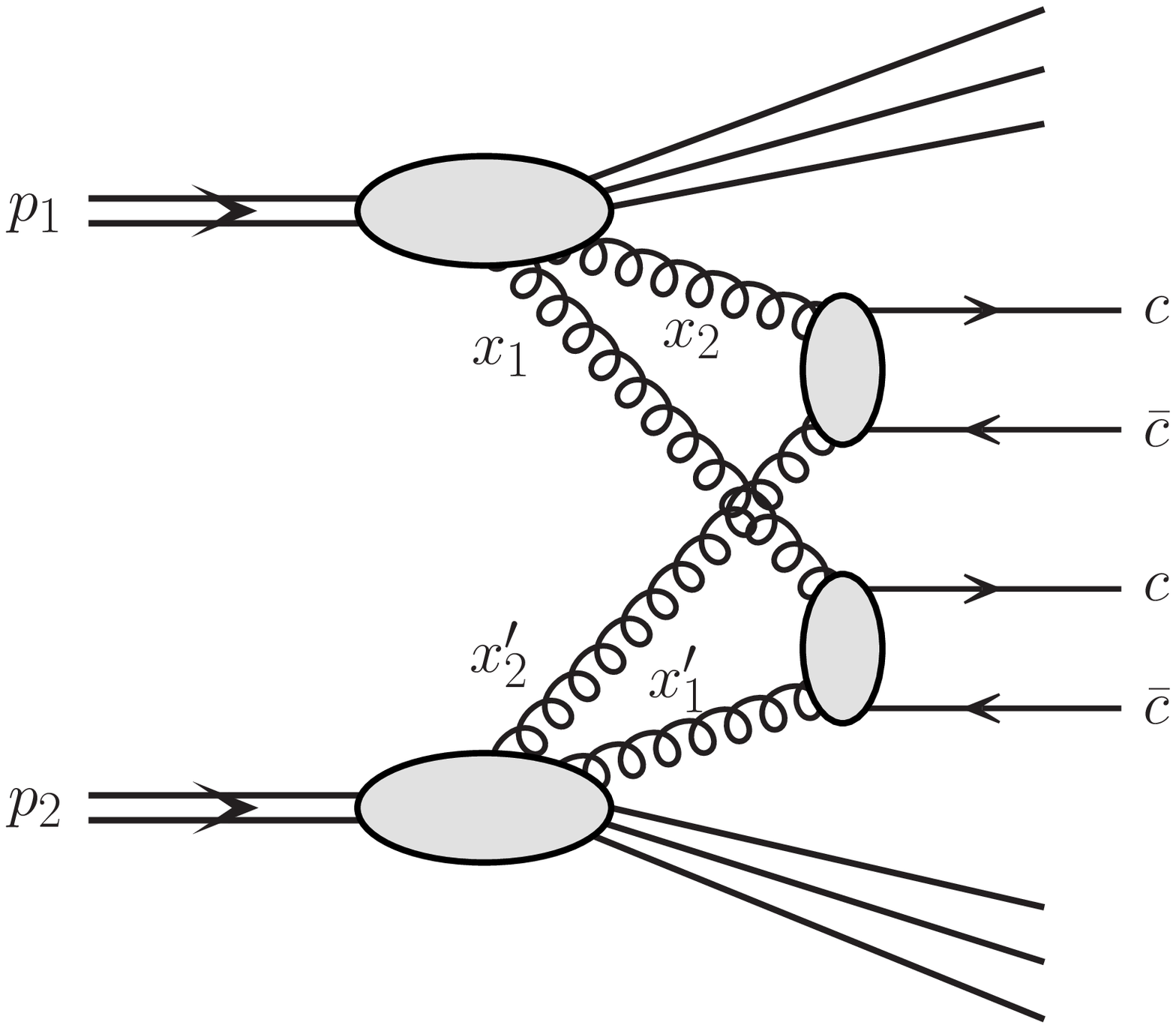}
\includegraphics[width=4cm]{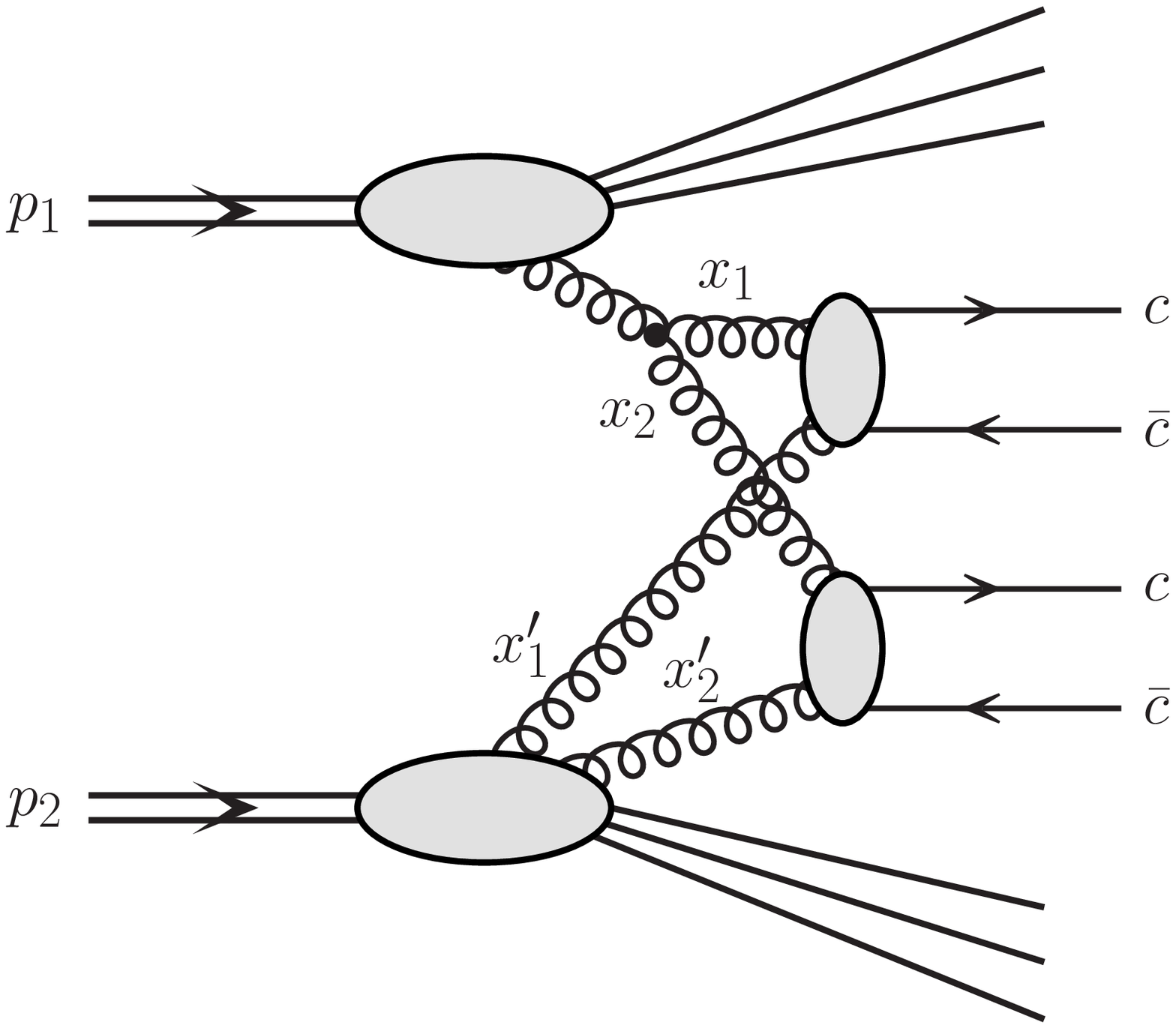}
\includegraphics[width=4cm]{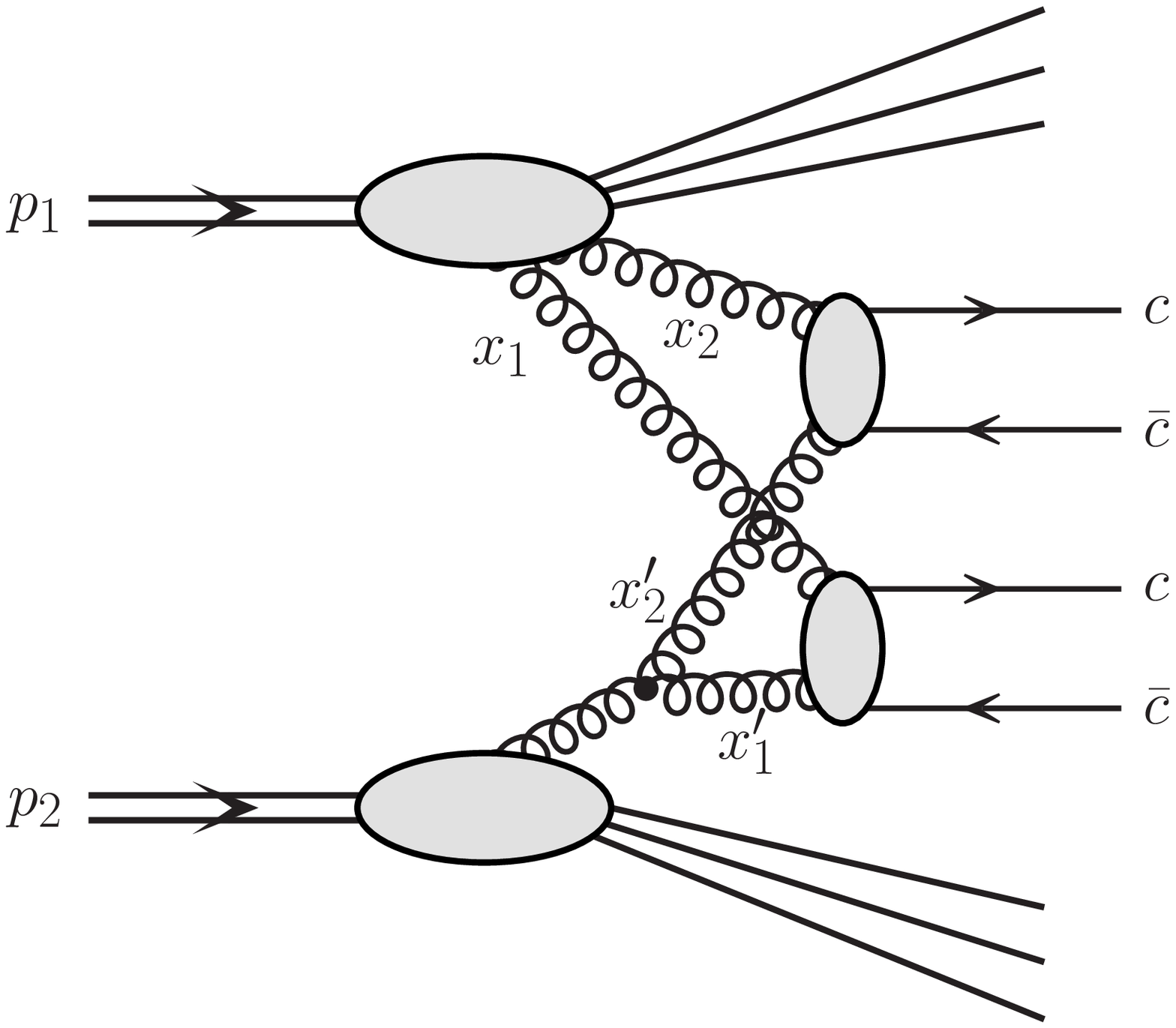}
\end{center}
   \caption{
\small The diagrams for DPS production of $c \bar c c \bar c$.
}
 \label{fig:diagrams_ccbarccbar}
\end{figure}

In Fig.~\ref{fig:diagrams_ccbarccbar} we illustrate a conventional
and perturbative DPS mechanisms
for $c \bar c c \bar c$ production. The 2v1 mechanism 
(the second and third diagrams) were considered first in \cite{GMS2014}.

In the case of $c \bar c c \bar c$ production the cross section for conventional DPS can be written as:
\begin{eqnarray}
\sigma(2v2) &=& \frac{1}{2} \frac{1}{\sigma_{eff,2v2}}
\int dy_1 dy_2 d^2 p_{1t} dy_3 dy_4 d^2 p_{2t} 
\frac{1}{16 \pi {\hat s}^2} \overline{ |{\cal M}(gg \to c \bar c)|^2} 
\; x_1 x_1' x_2 x_2' \nonumber \\ 
&& \times \;
D^{gg}(x_1, x_2, \mu_1^2, \mu_2^2) \;
D^{gg}(x_1, x_2, \mu_1^2, \mu_2^2)
\label{DPS_ccbarccbar_2v2}         
\end{eqnarray}
while that for the perturbative parton splitting DPS in a very
similar fashsion (see e.g.\cite{GMS2014})
\begin{eqnarray} 
\sigma(2v1) &=& \frac{1}{2} \frac{1}{\sigma_{eff,2v1}}
\int dy_1 dy_2 d^2 p_{1t} dy_3 dy_4 d^2 p_{2t}  
\frac{1}{16 \pi {\hat s}^2} \overline{ |{\cal M}(gg \to c \bar c)|^2} 
\; x_1 x_1' x_2 x_2' \nonumber \\
&& \!\!\!\!\!\!\!\!\!\!\!\!\!\!\!\!\!\!\!\!\!\!\!\!\!\!\!\!\!\!\! \times \; \left({\hat D}^{gg}(x_1', x_2', \mu_1^2, \mu_2^2) D^{gg}(x_1, x_2, \mu_1^2, \mu_2^2) 
+ \; D^{gg}(x_1', x_2', \mu_1^2, \mu_2^2) {\hat D}^{gg}(x_1, x_2, \mu_1^2, \mu_2^2)\right) \nonumber \\
\label{DPS_ccbarccbar_2v1}
\end{eqnarray}


\subsection{Four-jet production in DPS}

In the calculations performed in Ref.~\cite{MS2014} all partonic cross
sections are calculated only in leading order.
The cross section for dijet production can be written then as:
\begin{equation}
\frac{d \sigma(i j \to k l)}{d y_1 d y_2 d^2p_t} = \frac{1}{16 \pi^2 {\hat s}^2}
\sum_{i,j} x_1 f_i(x_1,\mu^2) \; x_2 f_j(x_2,\mu^2) \;
\overline{|\mathcal{M}_{i j \to k l}|^2} \;,
\label{LO_SPS}
\end{equation}
where $y_1$, $y_2$ are rapidities of the two jets and $p_t$ is
transverse momentum of one of them (identical).

In our calculations we include all leading-order $i j \to k l$ partonic 
subprocesses (see e.g. \cite{Ellis-Stirling-Webber,Barger-Phillips}).
The $K$-factor for dijet production is rather small, of the order of 
$1.1 - 1.3$ (see e.g. \cite{K-factor1,K-factor2}), 
and can be easily incorporated in our calculations. Below we shall show that
already the leading-order approach gives results in reasonable 
agreement with recent ATLAS \cite{ATLASjets} and CMS \cite{CMSjets} data.

This simplified leading-order approach was used
in our first estimate of DPS differential cross sections for jets 
widely separated in rapidity \cite{MS2014}. 
Similarly as for $c \bar c c \bar c$ production one can write:
\begin{eqnarray}
\frac{d \sigma^{DPS}(p p \to \textrm{4jets} \; X)}{d y_1 d y_2 d^{2} p_{1t} d y_3 d y_4 d^{2} p_{2t}} &=& \nonumber \\
&& \!\!\!\!\!\!\!\!\!\!\!\!\!\!\!\!\!\!\!\!\!\!\!\!\!\!\!\! \!\!\!\!\!\!\!\!\!\!\!\!\!\!\!\!\!\!\!\!\!\!\!\!\!\!\!\!  
 \sum\nolimits_{i_1,j_1,k_1,l_1,i_2,j_2,k_2,l_2} \; \frac{\mathcal{C}}{\sigma_{eff}} \;
\frac{d \sigma(i_1 j_1 \to k_1 l_1)}{d y_1 d y_2 d^{2} p_{1t}} \; \frac{d \sigma(i_2 j_2 \to k_2 l_2)}{d y_3 d y_4 d^{2} p_{2t}}\;,
\label{DPS}
\end{eqnarray}
where
$\mathcal{C} = \left\{ \begin{array}{ll}
\frac{1}{2}\;\; & \textrm{if} \;\;i_1 j_1 = i_2 j_2 \wedge k_1 l_1 = k_2 l_2\\
1\;\;           & \textrm{if} \;\;i_1 j_1 \neq i_2 j_2 \vee k_1 l_1 \neq k_2 l_2
\end{array} \right\} $ and partons $j,k,l,m = g, u, d, s, \bar u, \bar d, \bar s$. 
The combinatorial factors include identity of the two subprocesses.
Each step of the DPS is calculated in the leading-order approach (see
Eq.(\ref{LO_SPS})).
Above $y_1$, $y_2$ and $y_3$, $y_4$ are rapidities of partons in
first and second partonic subprocess, respectively.
The $p_{1t}$ and $p_{2t}$ are respective transverse momenta.

Experimental data from the Tevatron \cite{Tevatron} and the LHC 
\cite{Aaij:2011yc,Aaij:2012dz,Aad:2013bjm} 
provide an estimate of $\sigma_{eff}$ in the denominator of formula 
(\ref{DPS}). As in our recent paper \cite{Hameren2014} we take 
$\sigma_{eff}$ = 15 mb.
A detailed analysis of $\sigma_{eff}$ based on various experimental data
can be found in Refs.~\cite{Seymour:2013qka,Bahr:2013gkj}.

\subsection{$W^+ W^-$ production}

The diagram representating the double parton scattering
process is shown in Fig.~\ref{fig:DPS}.
The cross section for double parton scattering is often modelled
in the factorized anzatz which in our case would mean:

\begin{figure*}
\begin{center}
\includegraphics[width=5cm]{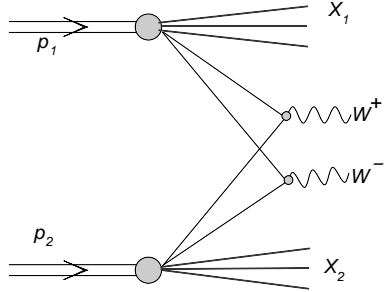}
\caption{Diagram representing double parton scattering mechanism
of production of $W^+ W^-$ pairs.
}
\label{fig:DPS}
\end{center}
\end{figure*}
%
\begin{equation}
\sigma_{W^+ W^-}^{DPS} = \frac{1}{\sigma_{qq}^{eff}} 
\sigma_{W^{+}}
\sigma_{W^{-}}
\; .
\label{factorized_model}
\end{equation}
In general, the parameter $\sigma_{q q}$ does not need to be the same
as for gluon-gluon initiated processes $\sigma_{gg}^{eff}$.
In the present, rather conservative, calculations we take it to be
$\sigma_{qq}^{eff} = \sigma_{gg}^{eff}$ = 15 mb.
The latter value is known within about 10 \% from 
systematics of gluon-gluon initiated processes at the Tevatron and LHC.

The factorized model (\ref{factorized_model}) can be generalized 
to more differential distributions.
For example in our case of $W^{+} W^{-}$ production the cross section
differential in $W$ boson rapidities can be written as:
\begin{equation}
\frac{d \sigma_{W^+ W^-}^{DPS}}{d y_{+} d y_{-}} =
\frac{1}{\sigma_{qq}^{eff}} 
\frac{d\sigma_W^{+}}{d y_{+}}
\frac{d\sigma_W^{-}}{d y_{-}} \; .
\label{generalized_factorized_model}
\end{equation}
In particular, in leading-order approximation the cross section for 
quark-antiquark annihilation reads:
\begin{eqnarray}
\frac{d\sigma}{dy} &=& \sum_{ij} 
\left( x_1 q_{i/1}(x_1,\mu^2) x_2 {\bar q}_{j/2}(x_2,\mu^2) 
+ x_1 {\bar q}_{i/1}(x_1,\mu^2) x_2 q_{j/2}(x_1,\mu^2) \right) \nonumber \\
&& \; \times \;  \overline{|{\cal M}_{ij \to W^{\pm}}|^2}
\label{rapidity_of_W}
\end{eqnarray}
where the matrix element for quark-antiquark annihilation to $W$ bosons
(${\cal M}_{ij \to W^{\pm}}$) contains Cabibbo-Kobayashi-Maskawa 
matrix elements.

When calculating the cross section for single $W$ boson production
in leading-order approximation a well known Drell-Yan $K$-factor can be 
included. The double-parton scattering would be then multiplied by $K^2$.

\section{Results for different processes}

\subsection{$c \bar c c \bar c$ production}

\begin{figure}[!h]
\begin{center}
\includegraphics[width=5.0cm]{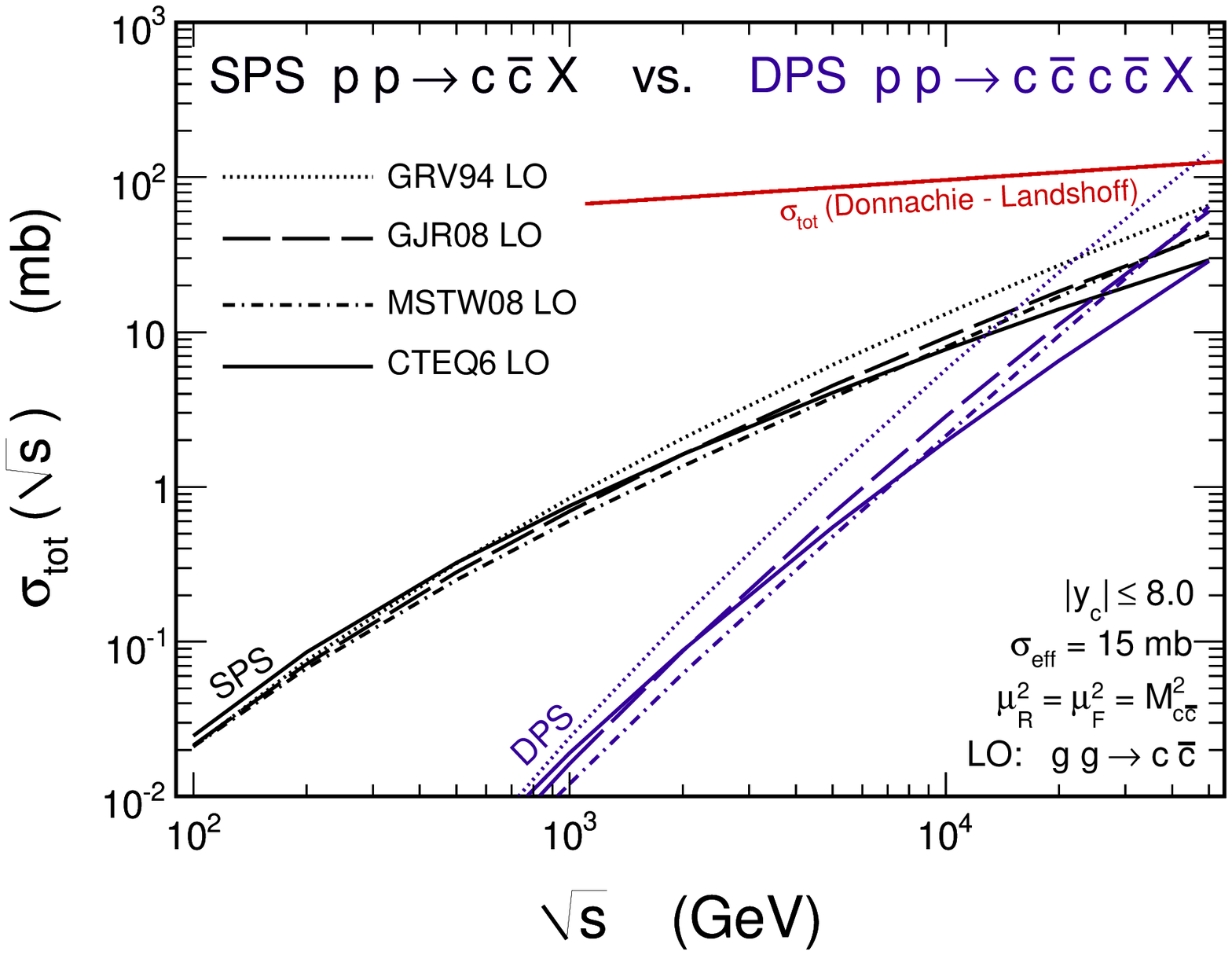}
\includegraphics[width=5.0cm]{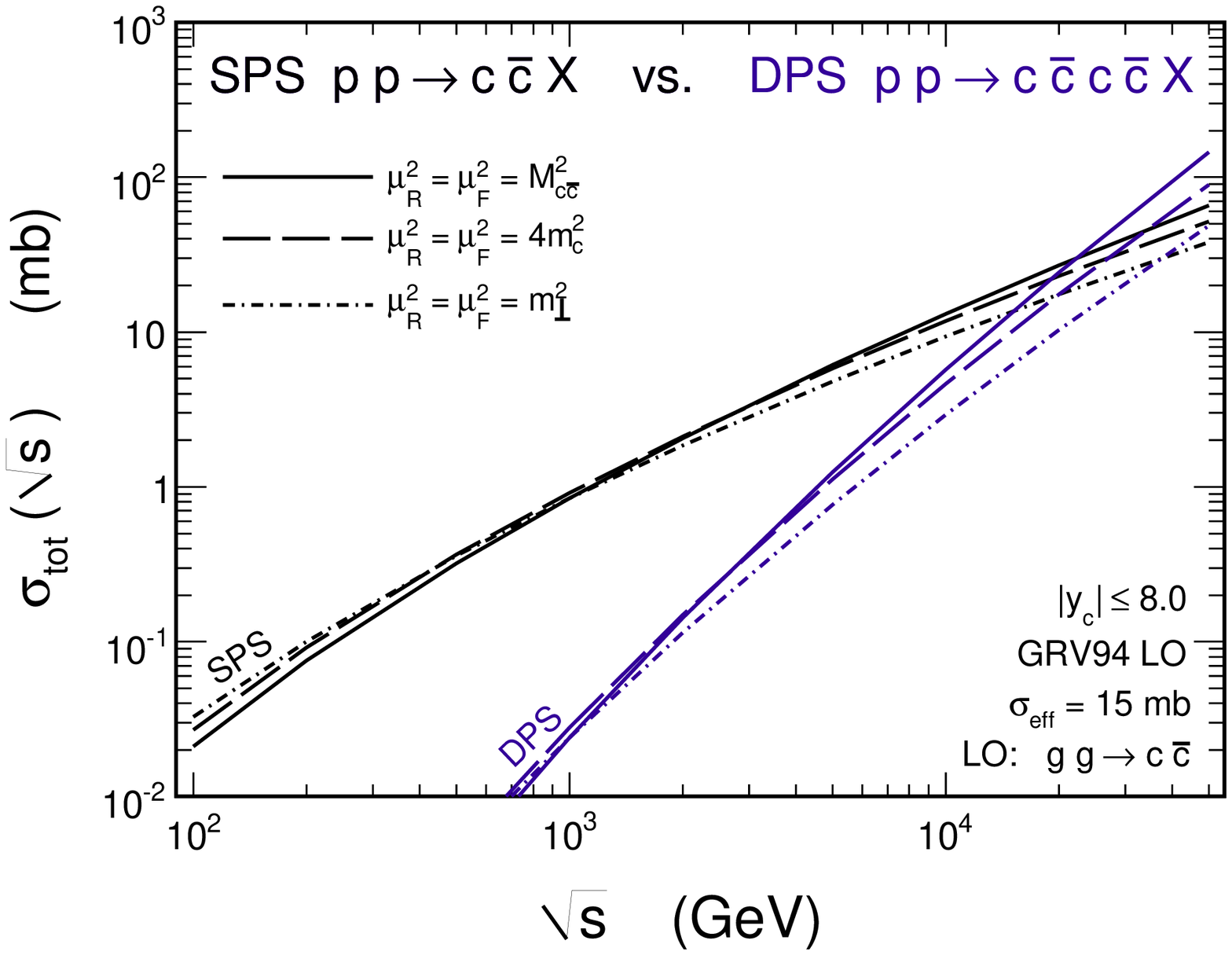}
\end{center}
\caption{
\small Total LO cross section for $c \bar c$ and double-parton
scattering production of $c \bar c c \bar c$ as a function of
center-of-mass energy (left panel) and uncertainties due to 
the choice of (factorization, renormalization) scales (right panel). 
We show in addition a parametrization of the total cross section 
in the left panel.
}
 \label{fig:single_vs_double_LO}
\end{figure}

We start presentation of our results with production of two pairs of $c \bar c$.
In Fig.~\ref{fig:single_vs_double_LO} we
compare cross sections for the single and double-parton
scattering as a function of proton-proton center-of-mass energy. 
At low energies the single-parton scattering dominates. For reference 
we show the proton-proton total cross section as a function
of collision energy as parametrized in Ref.~\cite{DL92}.
At low energy the $c \bar c$ or $ c \bar c c \bar c$ cross sections are much
smaller than the total cross section. At higher energies both the contributions
dangerously approach the expected total cross section.
This shows that inclusion of unitarity effect and/or saturation 
of parton distributions may be necessary.
The effects of saturation in $c \bar c$ production were included
e.g. in Ref.~\cite{enberg1}
but not checked versus experimental data. 
Presence of double-parton scattering changes the situation. The
double-parton scattering is potentially very important
ingredient in the context of high energy neutrino production
in the atmosphere \cite{GIT96, MRS2003, enberg1} or of cosmogenic origin
\cite{enberg2}.
At LHC energies the cross section for both terms become comparable.
This is a completely new situation.

\begin{figure}[!h]
\begin{center}
\includegraphics[width=7cm]{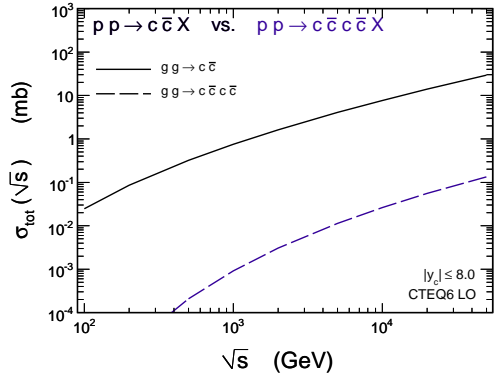}
\end{center}
\caption{Cross section for SPS production of $c \bar c c \bar c$
compared to this for standard $c \bar c$ production as a function
of collision energy.
}
\label{fig:SPSccbar_vs_SPSccbarccbar}
\end{figure}

So far we have concentrated on DPS production of $c \bar c c \bar c$
and completely ignored SPS production of $c \bar c c \bar c$. In
Refs.\cite{SS2012,Hameren2014} we calculated the SPS contribution
in high-energy approximation \cite{SS2012} and including all diagrams
in the collinear-factorization approach \cite{Hameren2014}.
In Fig.~\ref{fig:SPSccbar_vs_SPSccbarccbar} we show the cross section
from Ref.~\cite{Hameren2014}. The corresponding cross section at the LHC energies
is more than two orders of magnitude smaller than that for $c \bar c$
production i.e. much smaller than the DPS contribution discussed
in the previous figure.

\begin{figure}
\begin{center}
\includegraphics[width=6cm]{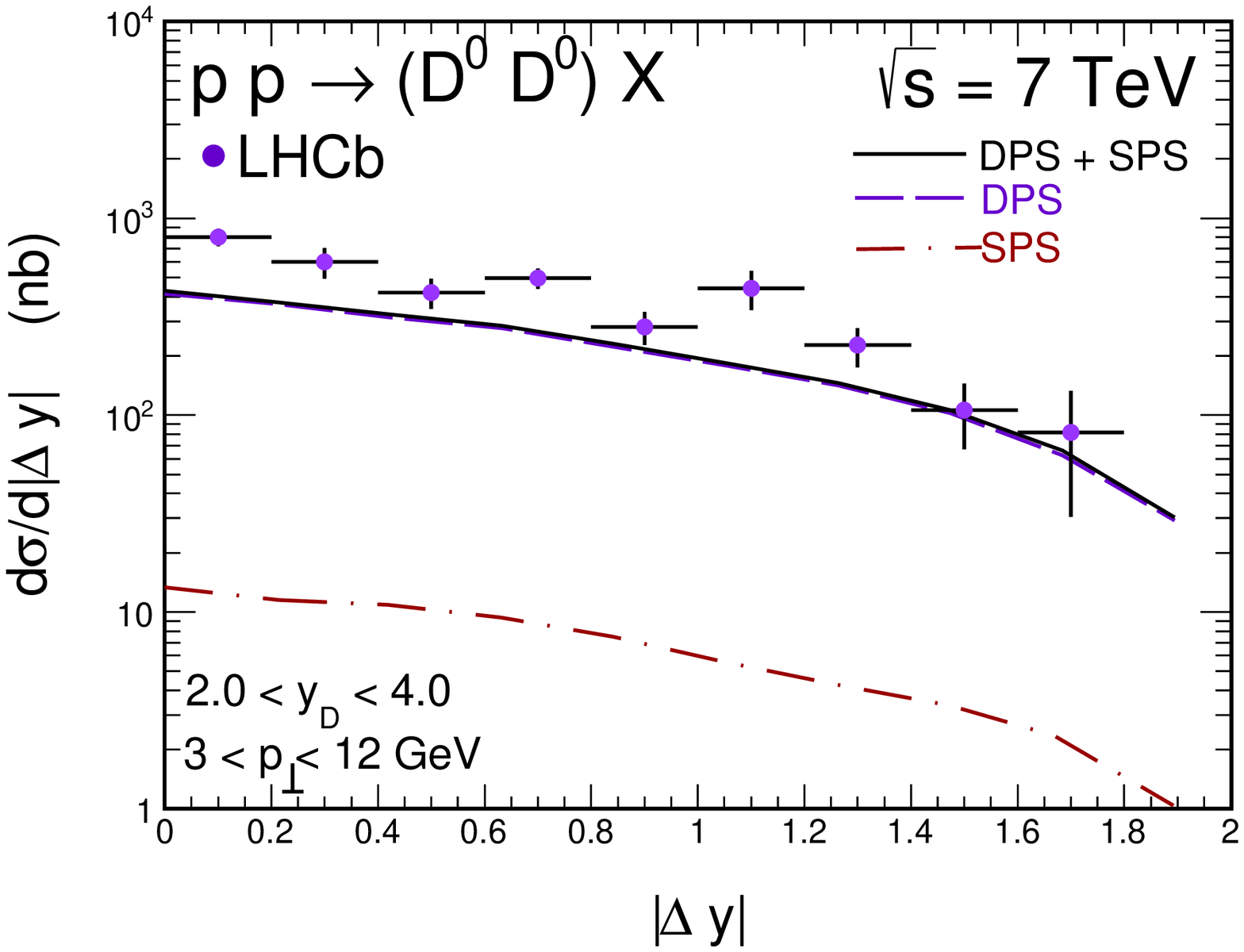}
\includegraphics[width=6cm]{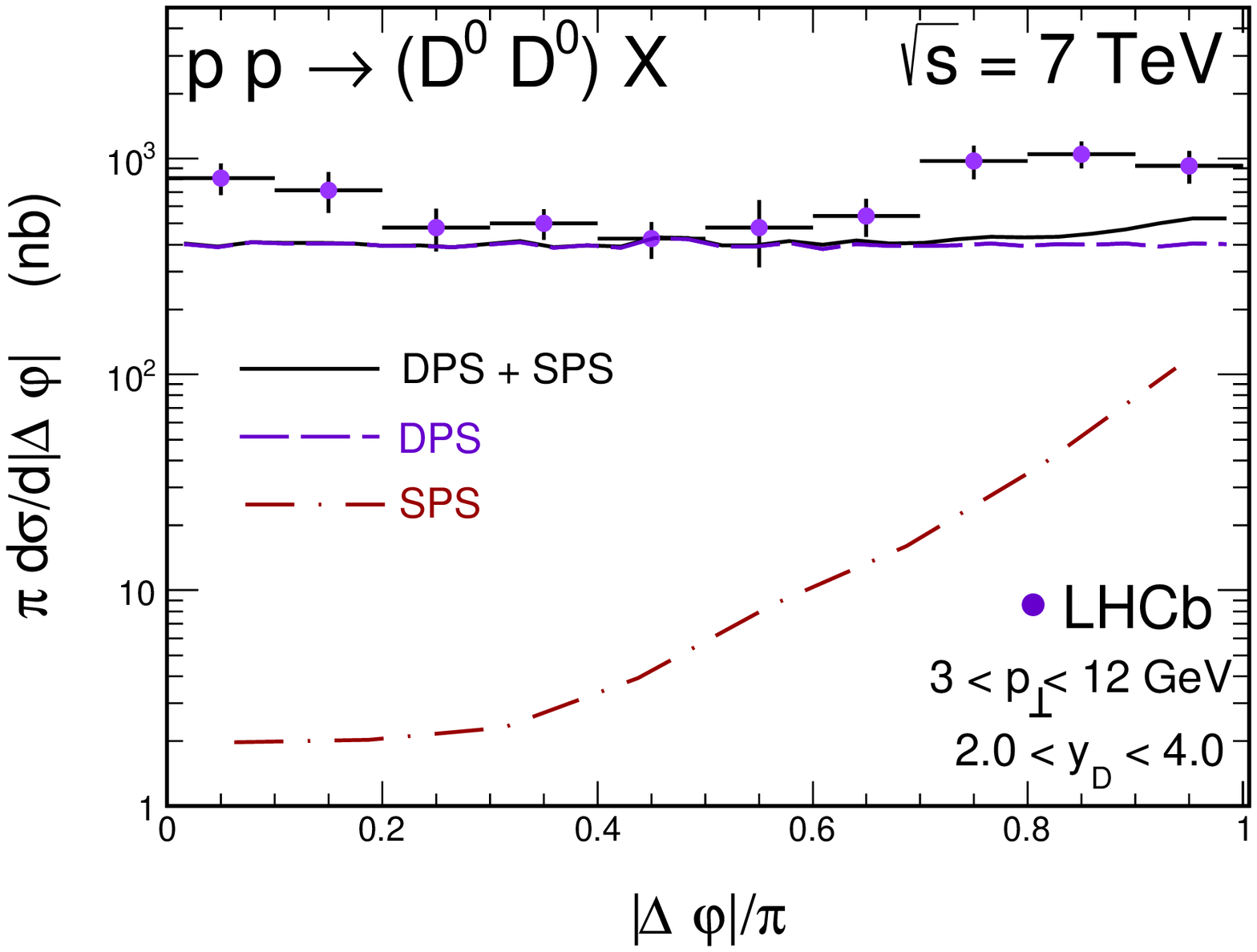}
\end{center}
\caption{Rapidity distance between two $D^0$ mesons (left panel) 
and corresponding azimuthal correlations (right panel).
}
\label{fig:DPS_ydiffandphid}
\end{figure}

In experiment one measures $D$ mesons instead of charm quarks/antiquarks.
In Fig.~\ref{fig:DPS_ydiffandphid} we show resulting distributions in
rapidity distance between two $D^0$ mesons (left panel) and
corresponding distribution in relative azimuthal angle (right
panel). The DPS contribuion (dashed line) dominates over the single
parton scattering one (dash-dotted line). The sum of the two
contributions is represented by the solid line. We get a reasonable
agreement with the LHCb experimental data \cite{Aaij:2012dz}.

\begin{figure}
\begin{center}
\includegraphics[width=6cm]{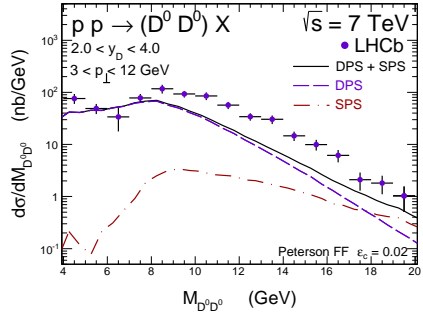}
\end{center}
\caption{Distribution of invariant mass of two $D^0$ mesons.}
\label{fig:dsig_dMinv}
\end{figure}

Distribution in the invariant mass of two $D^0 D^0$ mesons is
shown in Fig.~\ref{fig:dsig_dMinv}. Again a reasonable agreement is
obtained. Some strength is missing in the interval 10 GeV $< M_{D^0 D^0} <$ 16 GeV.

\begin{figure}
\begin{center}
\includegraphics[width=5cm]{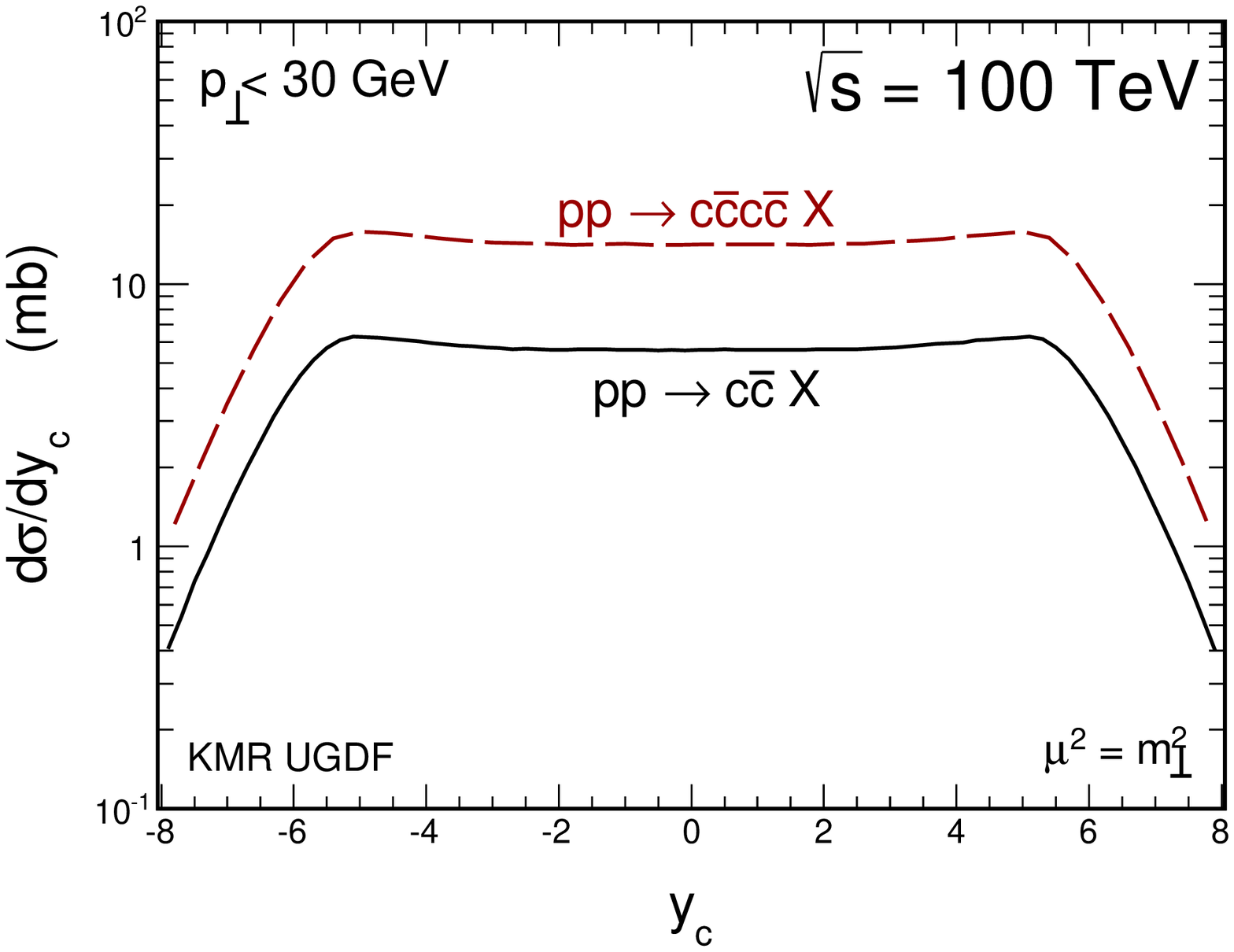}
\includegraphics[width=5cm]{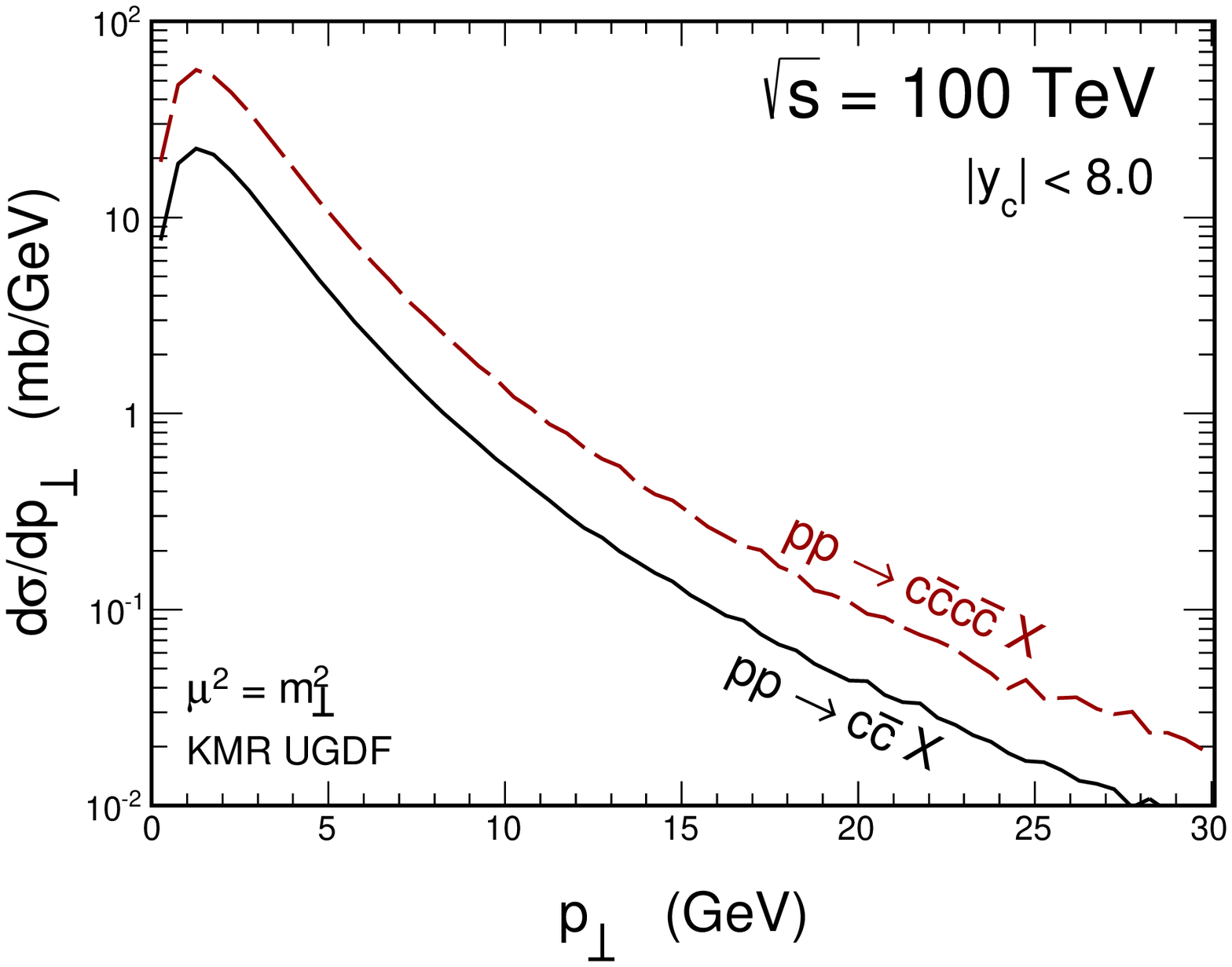}
\end{center}
\caption{Cross section for one $c$ or one $\bar c$ from the 
$c \bar c c \bar c$ final state at the FCC.}
\label{fig:charm_FCC}
\end{figure}

At the LHC the cross section for $p p \to c \bar c$ is still bigger
than that for $p p \to c \bar c c \bar c$ \cite{MS2013}.
As shown in Fig.~\ref{fig:single_vs_double_LO} the latter cross section
is growing fast and at high energy it may become even larger than that for
single pair production. The situation at Future Circular Collider (FCC) 
is shown in Fig.~\ref{fig:charm_FCC}. Now the situation reverses
and the cross section for $c \bar c c \bar c$ is bigger than that
for single pair production. We predict rather flat distributions
in charm quark/antiquark rapidities. The shapes in quark/aniquark
transverse momenta are almost identical which can be easily understood
within the formalism presented in the previous section.

\subsection{Parton splitting}

As described in the Formalism section the splitting contributions are
calculated in leading order only.
In the calculations performed in Ref.~\cite{GMS2014} we either assumed 
$\mu_1^2 = m_{1t}^2$ and $\mu_2^2 = m_{2t}^2$,
or $\mu_1^2 = M_{c\bar c,1}^{2}$ and $\mu_2^2 = M_{c\bar c,2}^{2}$.
The quantity $m_{it}$ is the transverse mass of either parton emerging
from subprocess $i$, whilst $M_{c \bar c,i}$ is the invariant mass of 
the pair emerging from subprocess $i$.

\begin{figure}
\begin{center}
\includegraphics[width=7cm]{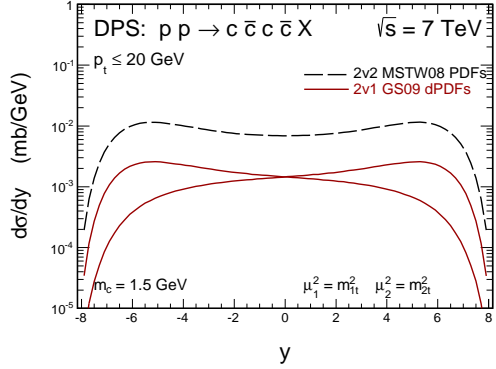}
\end{center}
\caption{ 
Rapidity distribution of charm quark/antiquark 
for $\sqrt{s}$ = 7 TeV for $\mu_1^2 = m_{1t}^2$, $\mu_2^2 = m_{2t}^2$.
}
\label{fig:dsig_dy}
\end{figure}

In Fig.~\ref{fig:dsig_dy} we show the rapidity distribution of the charm
quark/antiquark for different choices of the scale at $\sqrt{s}$ = 7 TeV. 
The conventional and splitting terms are shown separately. 
The splitting contribution (lowest curve, red online) is smaller, 
but has almost the same shape as the conventional DPS contribution. 
We can observe asymmetric (in rapidity) shapes for the $1v2$ and $2v1$ contributions.

\begin{figure}
\begin{center}
\includegraphics[width=6cm]{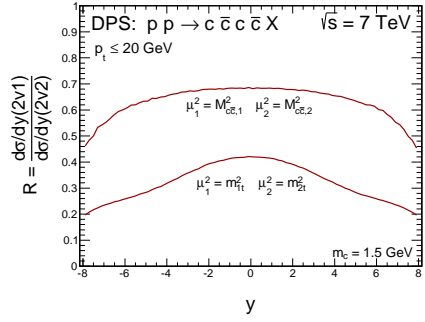}
\end{center}
\caption{
Ratio of the 2v1 to 2v2 cross sections as a function of quark/antiquark
rapidity.}
\label{fig:ratio_y_charm_2v1vs2v2}
\end{figure}

The corresponding ratios of the 2v1-to-2v2 contributions 
as a function of rapidity is shown in 
Fig.~\ref{fig:ratio_y_charm_2v1vs2v2}. 

\begin{figure}
\begin{center}
\includegraphics[width=6cm]{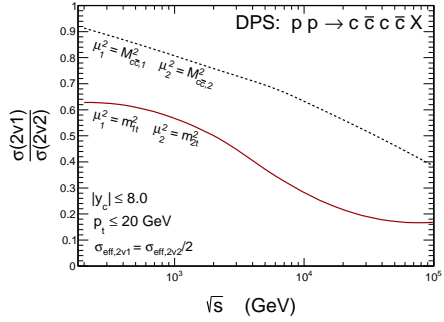}
\end{center}
\caption{
Ratio of the 2v1 to 2v2 cross sections as a function of collision energy.}
\label{fig:ratio_sqrtS_charm}
\end{figure}

In Fig.~\ref{fig:ratio_sqrtS_charm} we show energy deprendence
of the ratio of the 2v1 to 2v2 cross sections. The ratio systematically
decreases with the collision energy.

\begin{figure}
\begin{center}
\includegraphics[width=6cm]{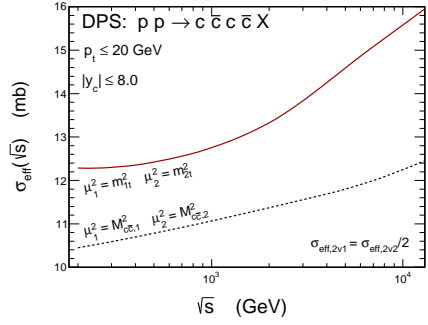}
\end{center}
\caption{
Energy and factorization scale dependence of $\sigma_{eff}$
for $c \bar c c \bar c$ production as a consequence of existence
of the two DPS components. 
In this calculation
$\sigma_{eff,2v2}$ = 30 mb and $\sigma_{eff,2v1}$ = 15 mb.
}
\label{fig:sig_eff_charm}
\end{figure}

Finally in Fig.~\ref{fig:sig_eff_charm} we show the empirical $\sigma_{eff}$,
for double charm production. Again $\sigma_{eff}$ 
rises with the centre-of-mass energy. A sizeable  difference of results for
different choices of scales can be observed.

\subsection{Jets with large rapidity separation}

\begin{figure}[!h]
\begin{center}
\includegraphics[width=4cm]{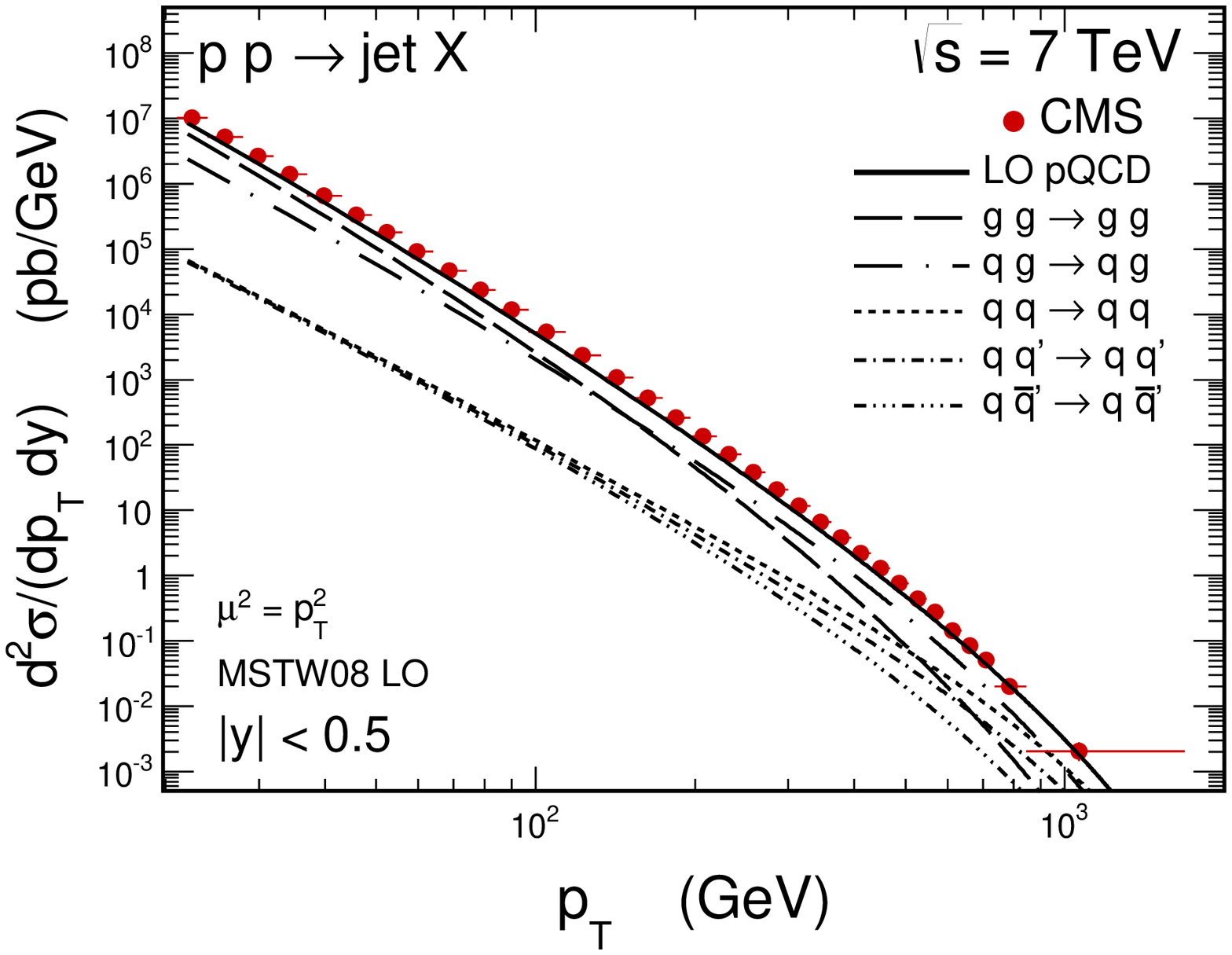}
\includegraphics[width=4cm]{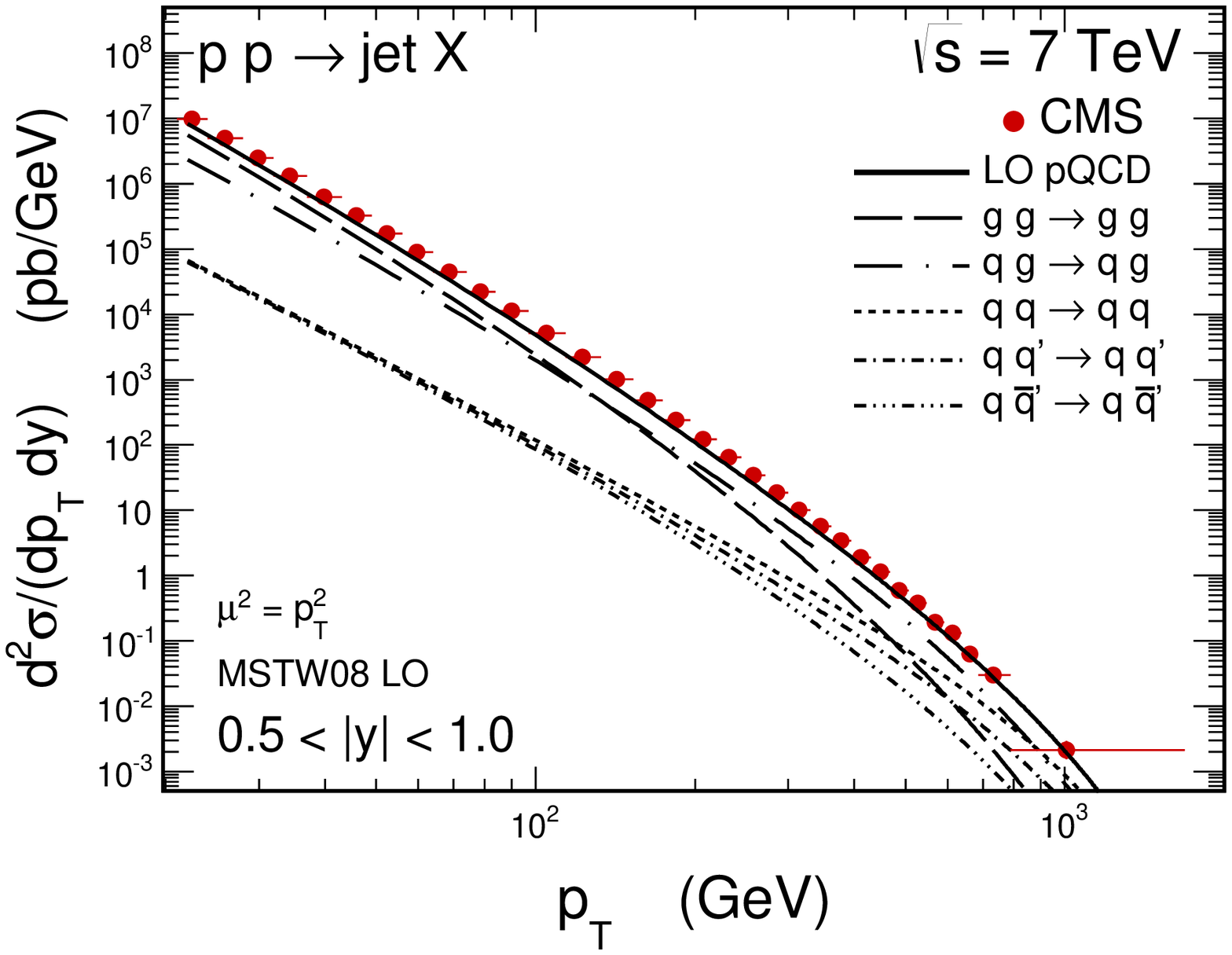}
\includegraphics[width=4cm]{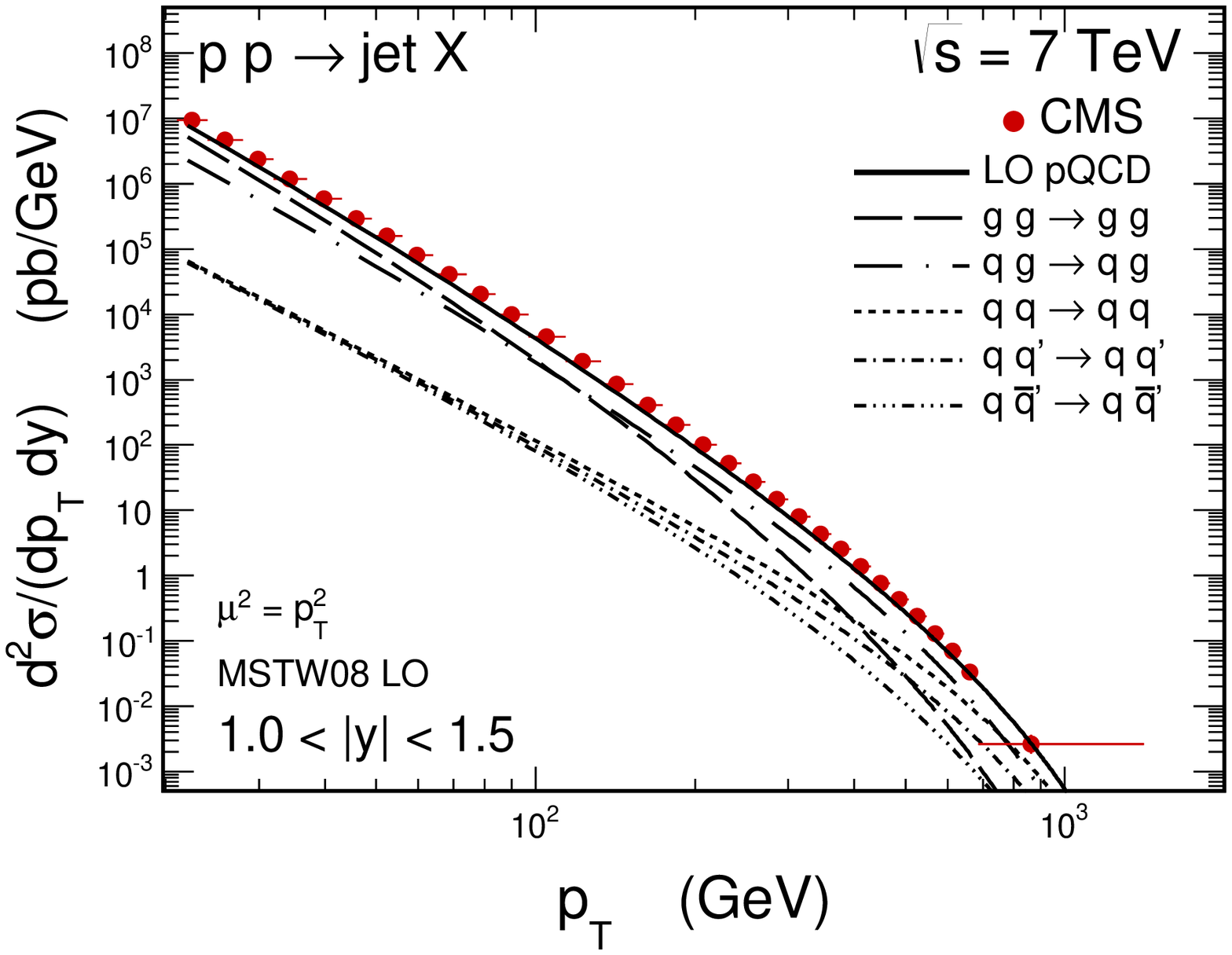}
\includegraphics[width=4cm]{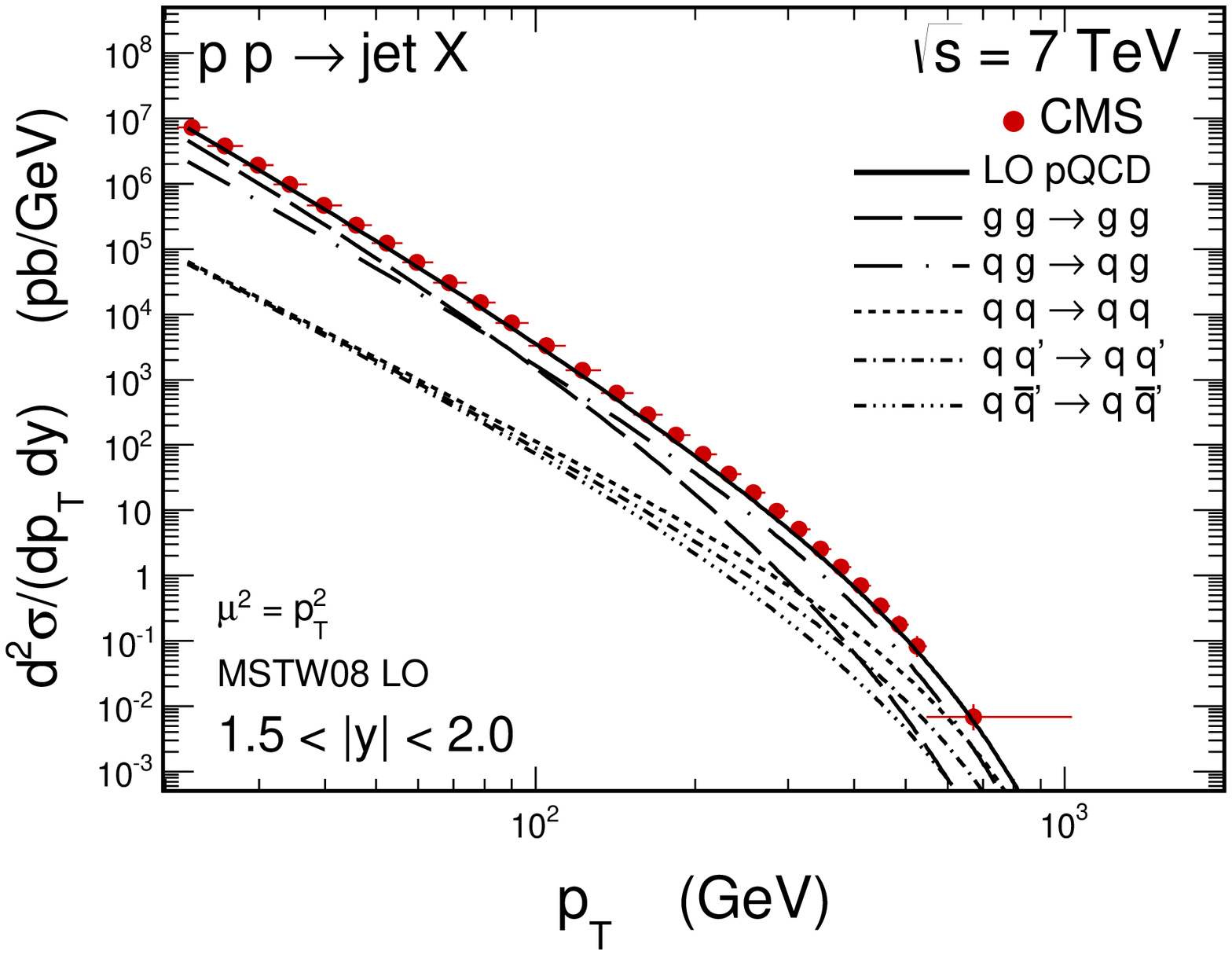}
\includegraphics[width=4cm]{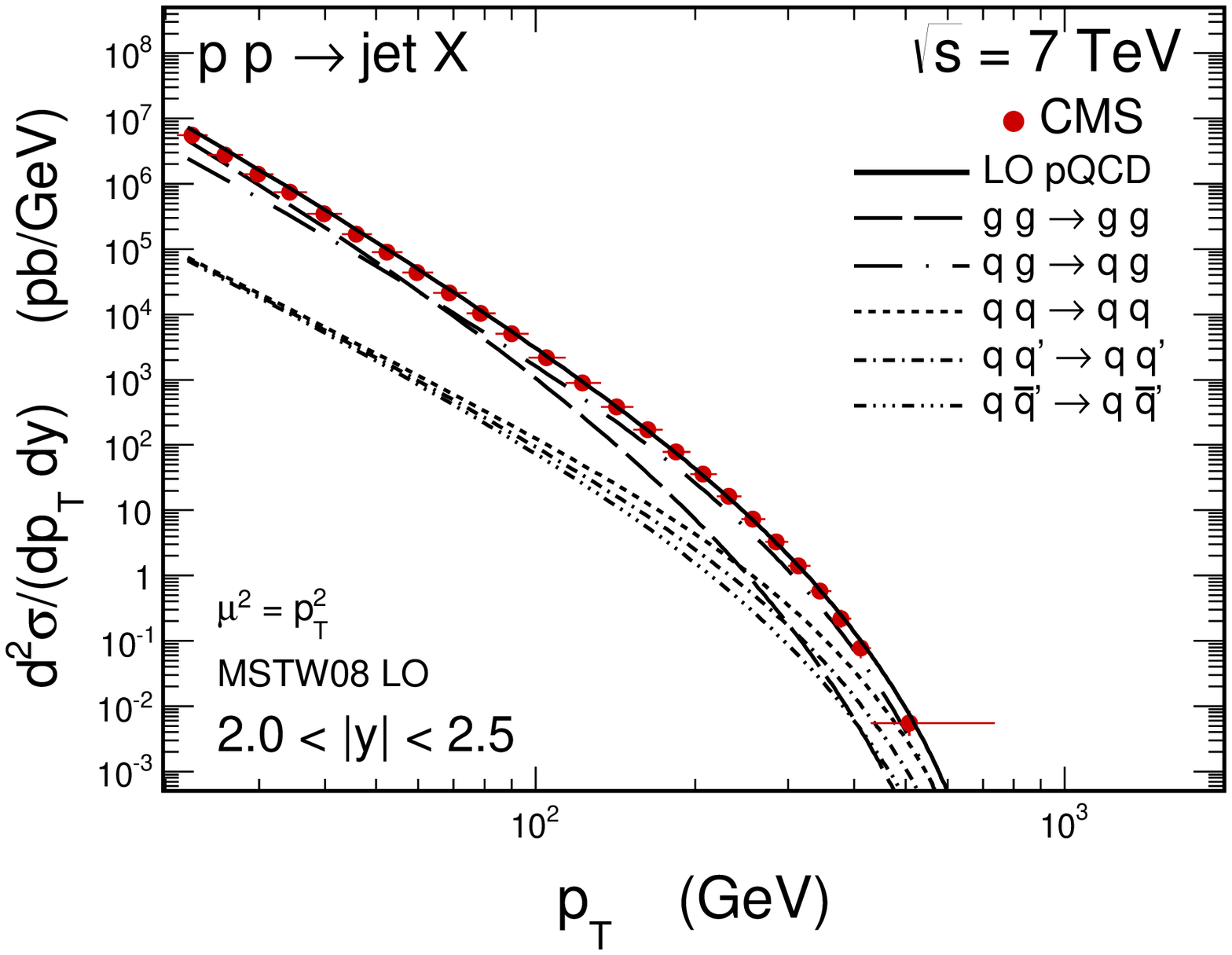}
\includegraphics[width=4cm]{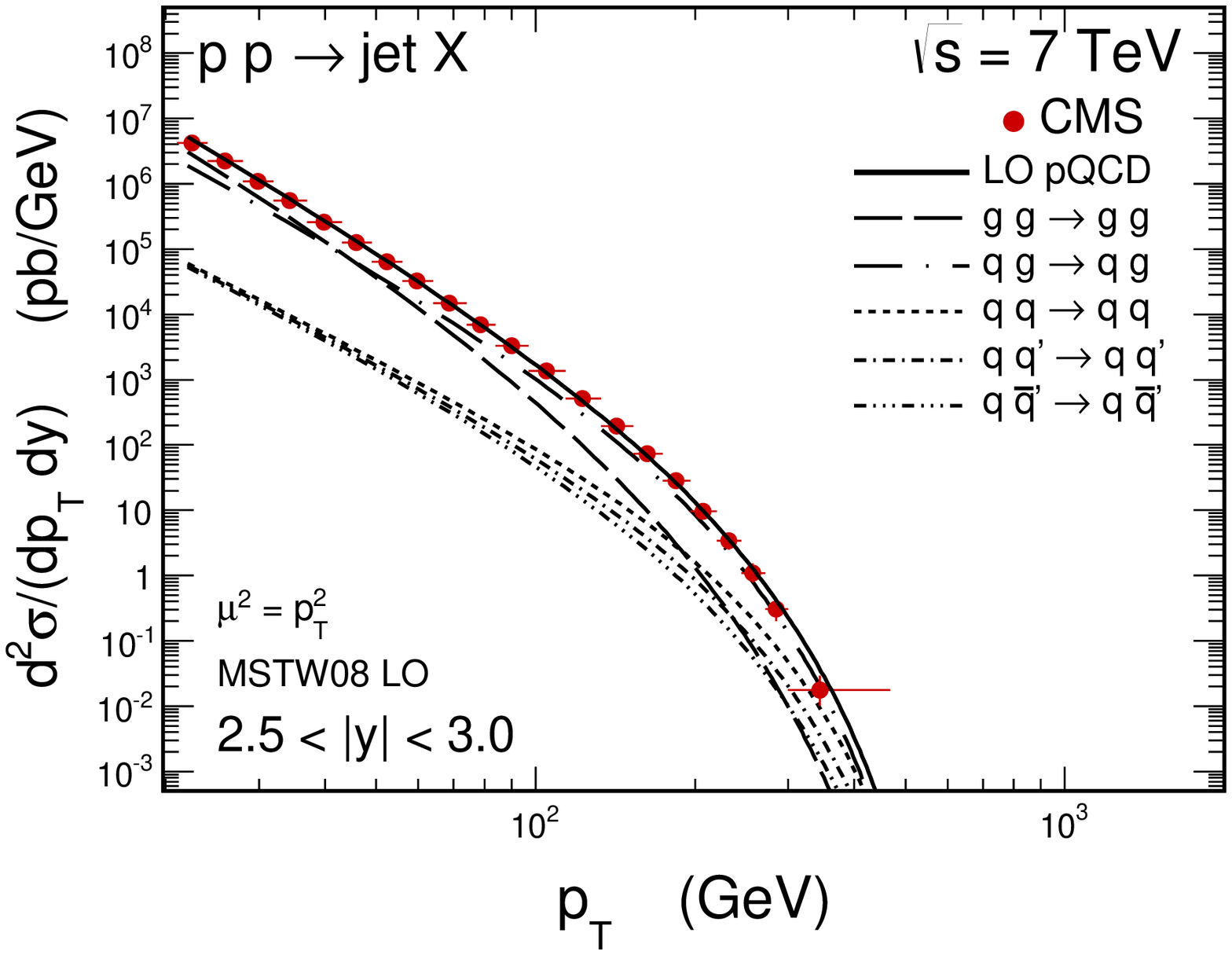}
\end{center}
   \caption{
\small Our results for inclusive jet production against the CMS
experimental data \cite{CMSjets}.
In addition we show decomposition into different partonic components 
as explained in the figure caption.
}
 \label{fig:pt-and-y-spectra-CMSjets}
\end{figure}

In Fig.~\ref{fig:pt-and-y-spectra-CMSjets} we compare our calculation for inclusive jet production with 
the CMS data \cite{CMSjets}. In addition, we show contributions of different
partonic mechanisms. In all rapidity intervals the gluon-gluon and quark-gluon
(gluon-quark) contributions clearly dominate over the other
contributions and in practice it is sufficient to include only
these subprocesses in further analysis.

\begin{figure}[!h]
\begin{center}
\includegraphics[width=5cm]{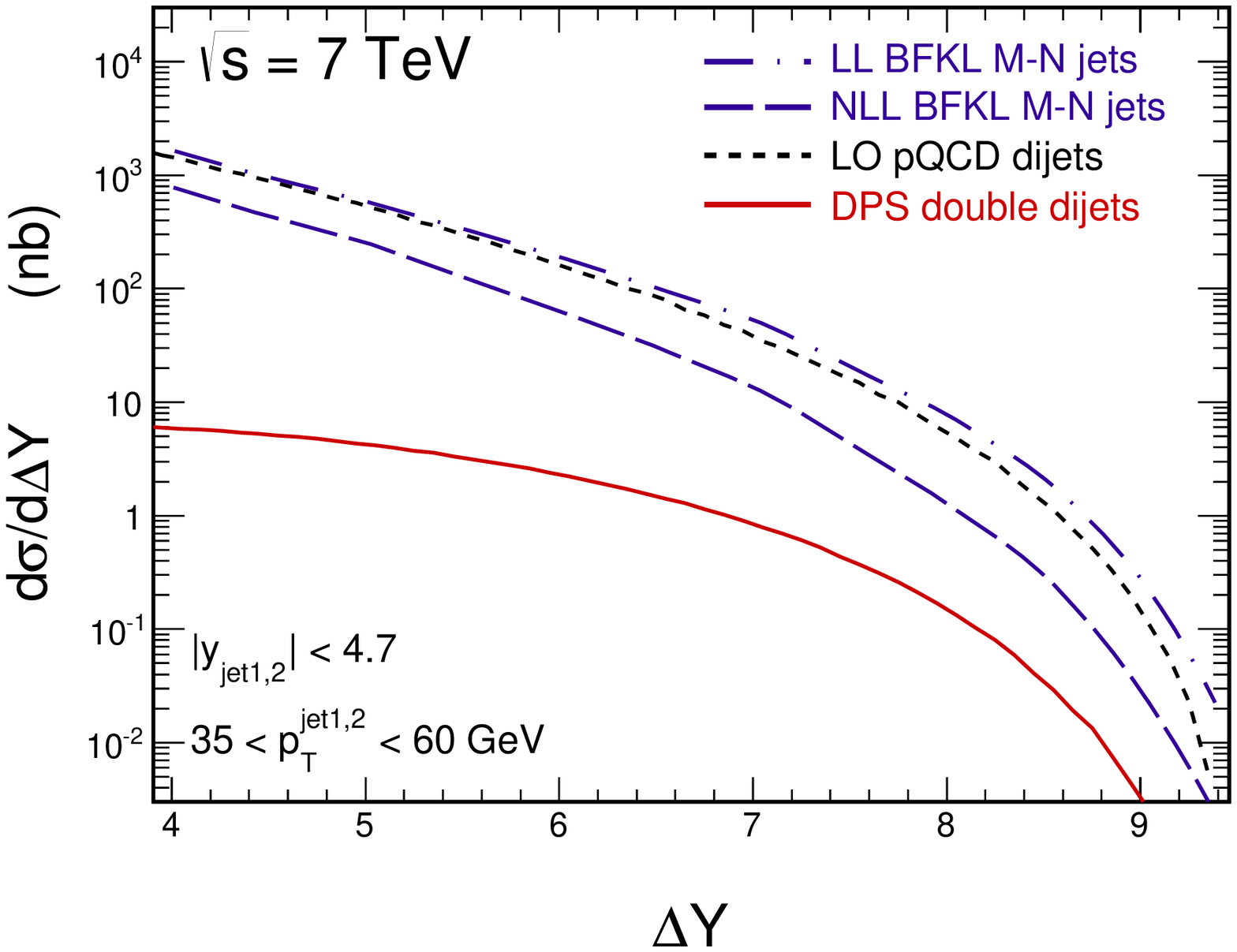}
\includegraphics[width=5cm]{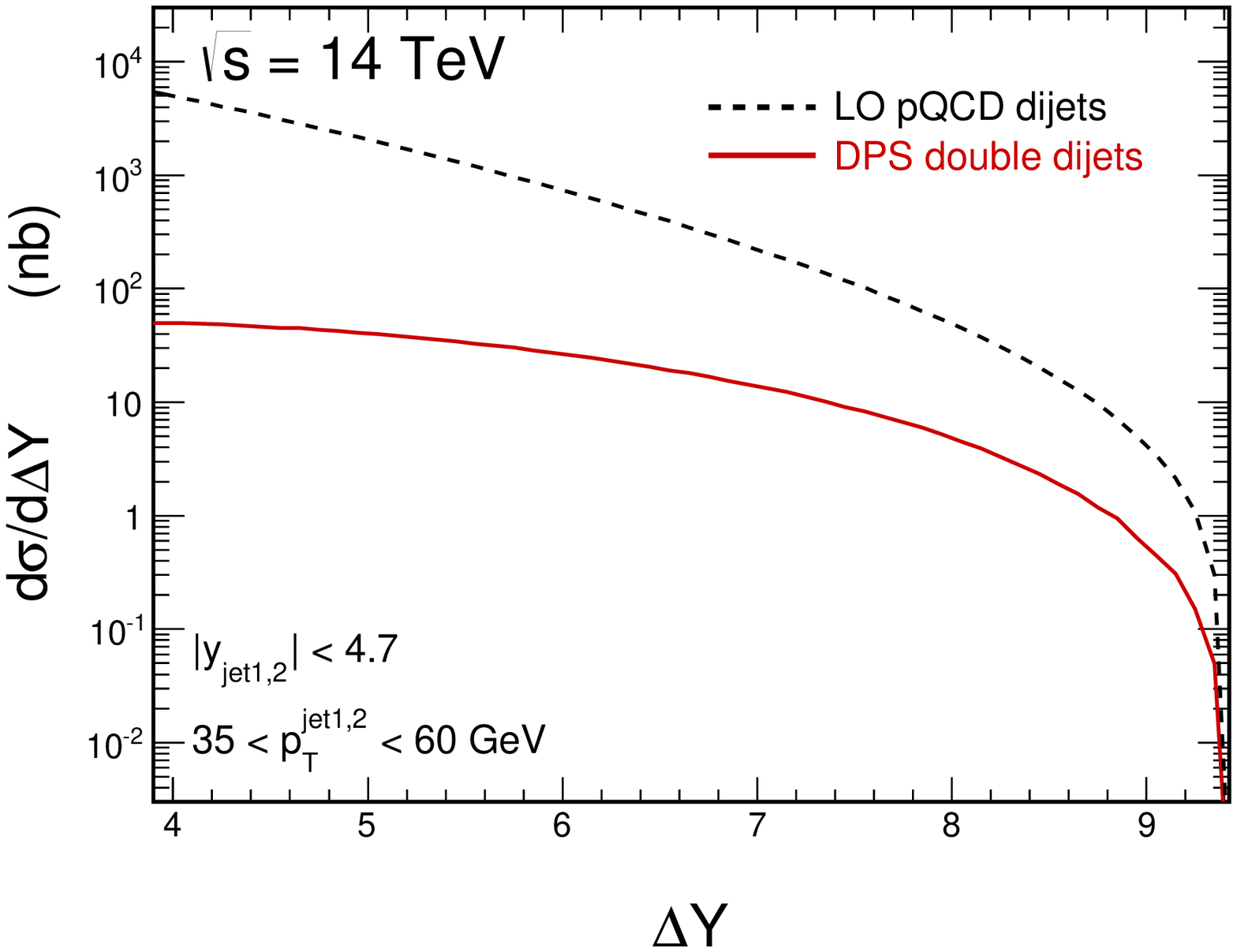}
\end{center}
   \caption{
\small Distribution in rapidity distance between jets 
(35 GeV $< p_t <$ 60 GeV) 
with maximal (the most positive) and minimal (the most negative)
rapidities. The collinear pQCD result is shown by the short-dashed line
and the DPS result by the solid line for $\sqrt{s}$ = 7 TeV (left panel)
and $\sqrt{s}$ = 14 TeV (right panel). For comparison we show also
results for the BFKL Mueller-Navelet jets in leading-logarithm 
and next-to-leading-order logarithm approaches from 
Ref.~\cite{Ducloue:2013hia}.
}
 \label{fig:Deltay1}
\end{figure}

Now we proceed to the jets with large rapidity separation.
In Fig.~\ref{fig:Deltay1} we show distribution in the rapidity 
distance between two jets in leading-order collinear calculation
and between the most distant jets in rapidity in the case of four DPS jets.
In this calculation we have included cuts for the
CMS expriment \cite{CMS_private}: $y_1, y_2 \in$ (-4.7,4.7),
$p_{1t}, p_{2t} \in$ (35 GeV, 60 GeV).
For comparison we show also results for the BFKL calculation from
Ref.~\cite{Ducloue:2013hia}. For this kinematics the DPS jets
give sizeable contribution only at large rapidity distance.
The NLL BFKL cross section (long-dashed line) is smaller than that for 
the LO collinear approach (short-dashed line).

\begin{figure}[!h]
\begin{center}
\includegraphics[width=5cm]{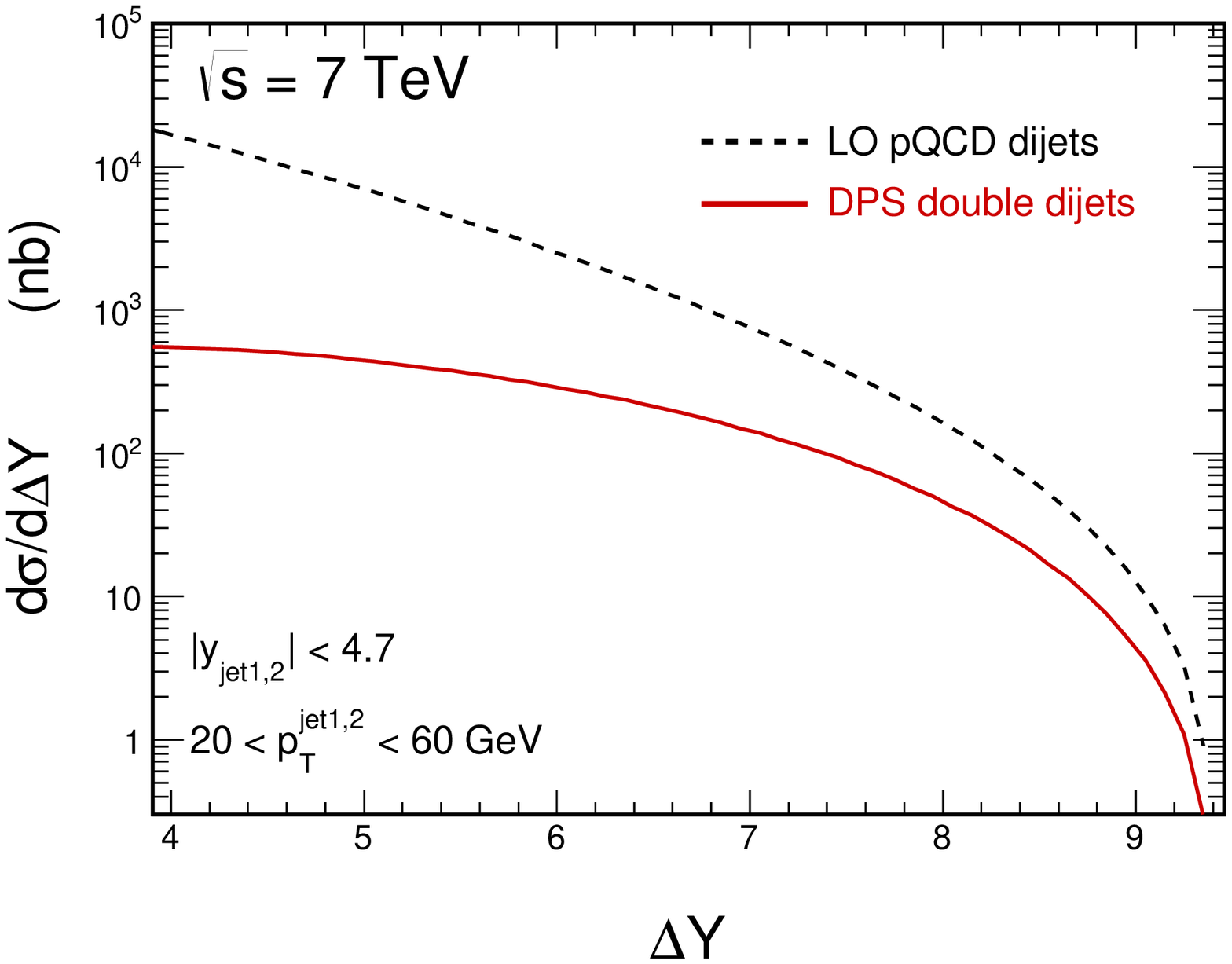}
\includegraphics[width=5cm]{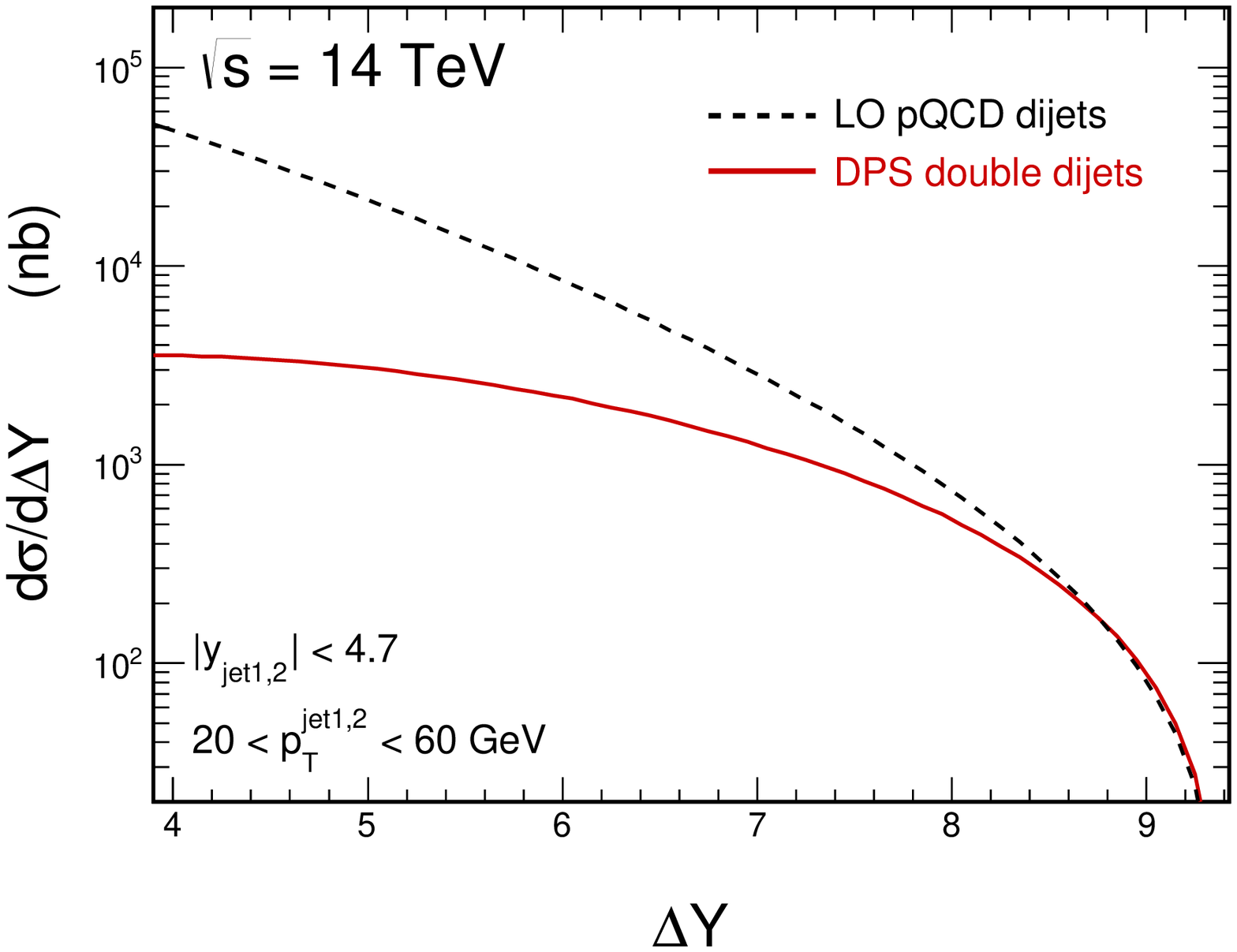} \\
\includegraphics[width=5cm]{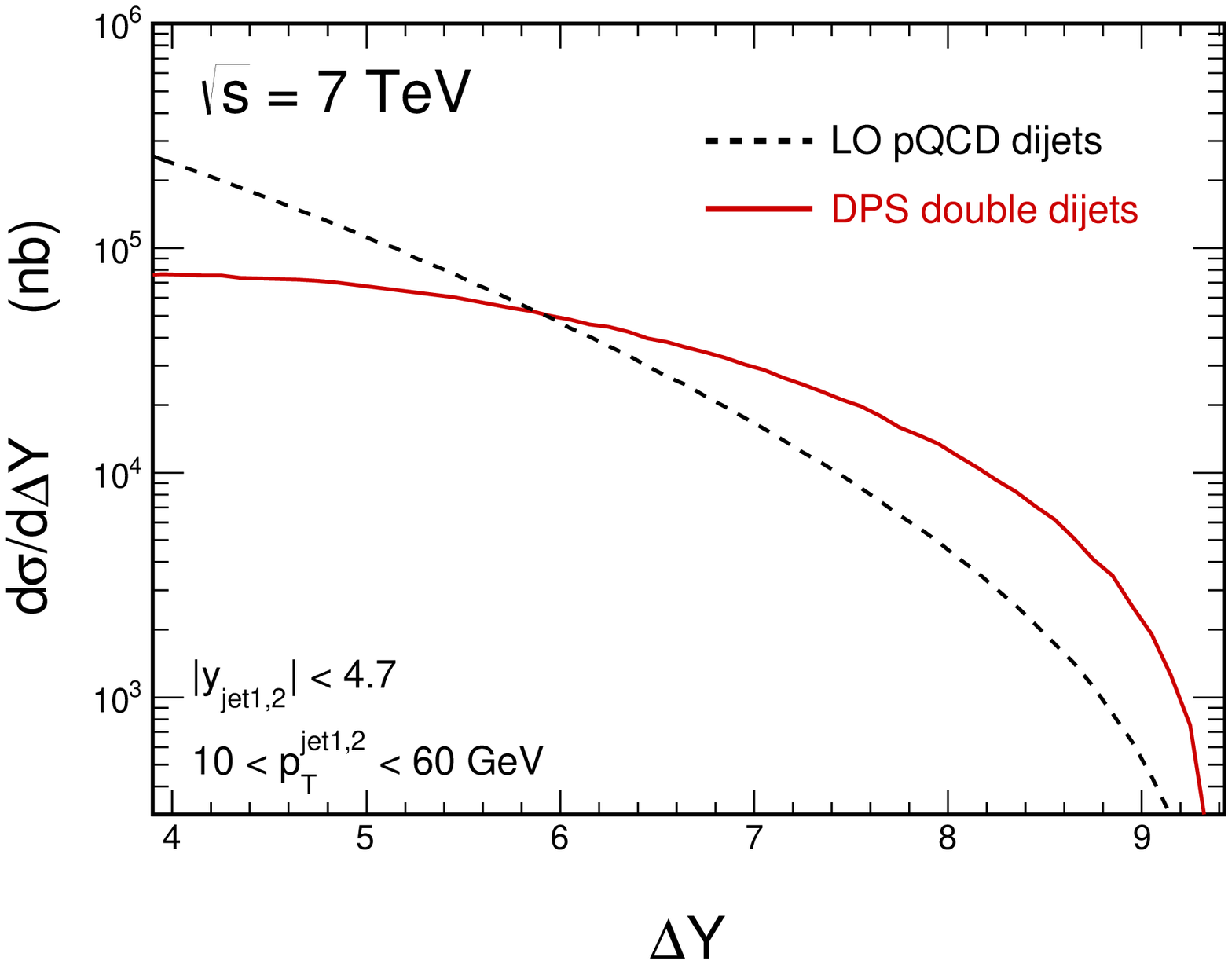}
\includegraphics[width=5cm]{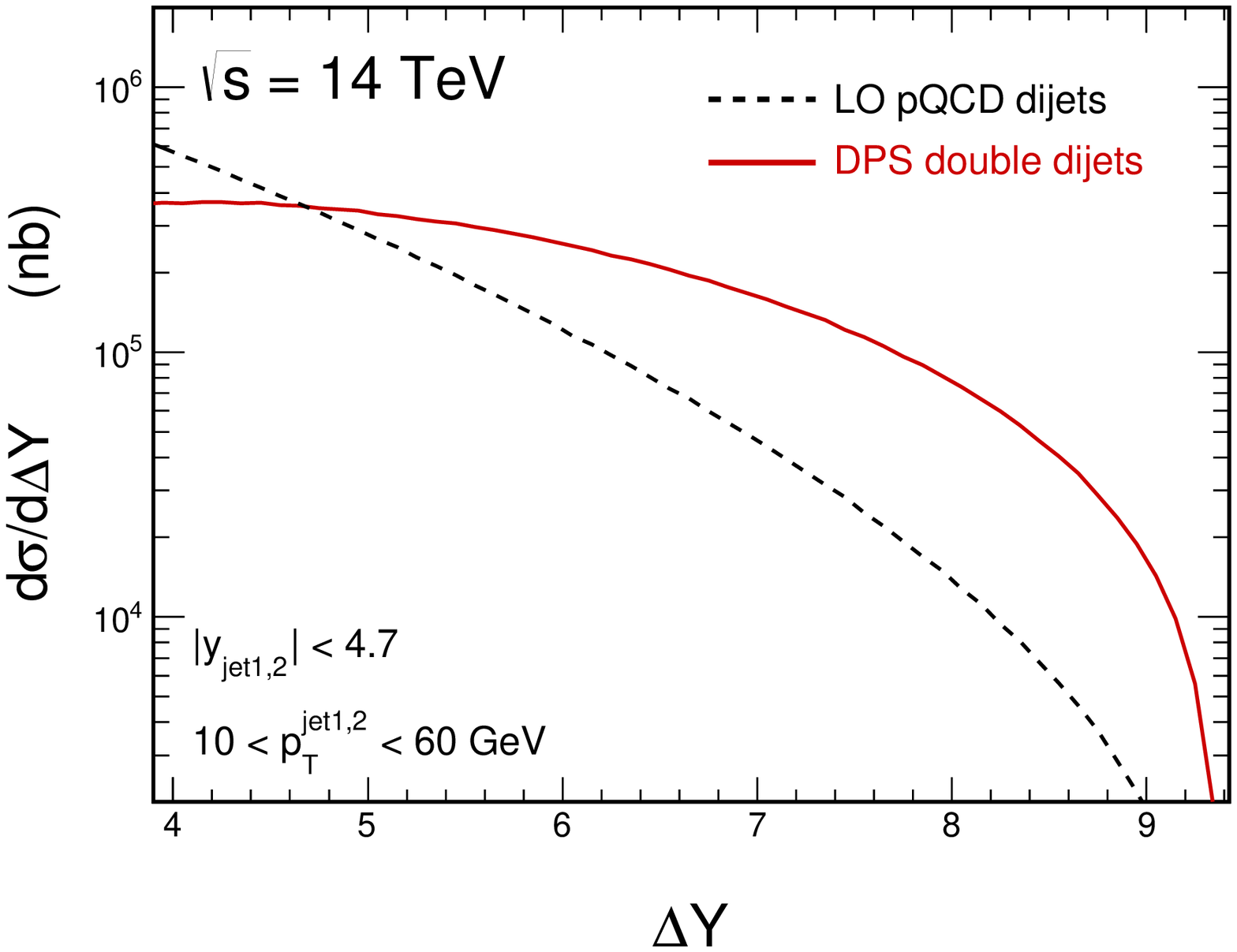}
\end{center}
 \caption{
\small The same as in the previous figure but now for somewhat smaller 
lower cut on minijet transverse momentum.
}
 \label{fig:Deltay-2}
\end{figure}

In Fig.~\ref{fig:Deltay-2} we show rapidity-distance
distribution for even smaller lowest transverse momenta of 
the "jet". A measurement of such minijets may be, however, difficult. 
Now the DPS contribution may even exceed the standard SPS 
dijet contribution, especially at the nominal LHC energy. 
How to measure such (mini)jets is an open issue. In principle,
one could measure correlations of 
semihard ($p_t \sim$ 10 GeV) neutral pions with the help of 
so-called zero-degree calorimeters (ZDC) which are installed by
all major LHC experiments.

\subsection{Production of $W^+ W^-$ pairs}

\begin{figure}
\begin{center}
\includegraphics[width=4.1cm]{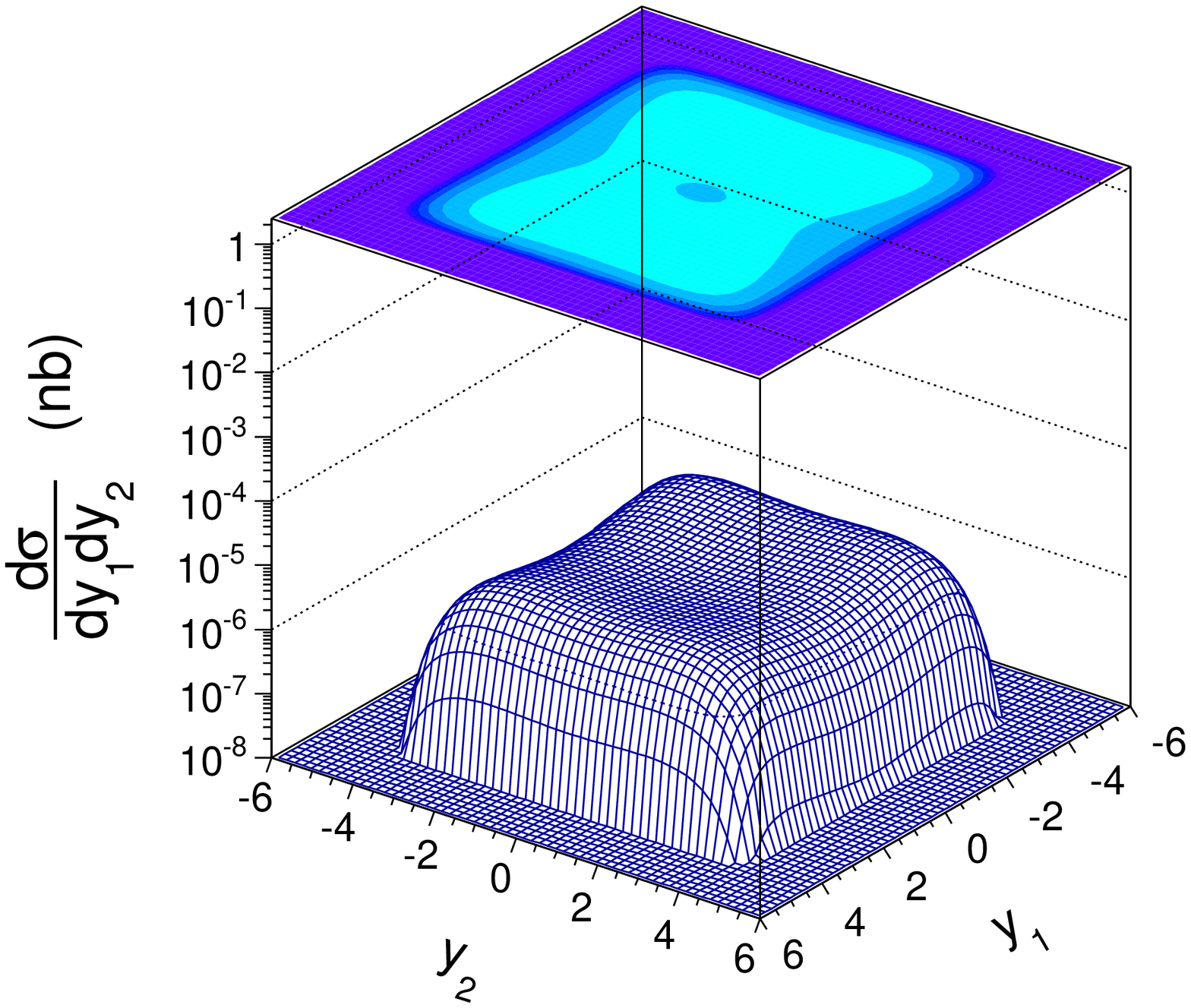}
\includegraphics[width=4.1cm]{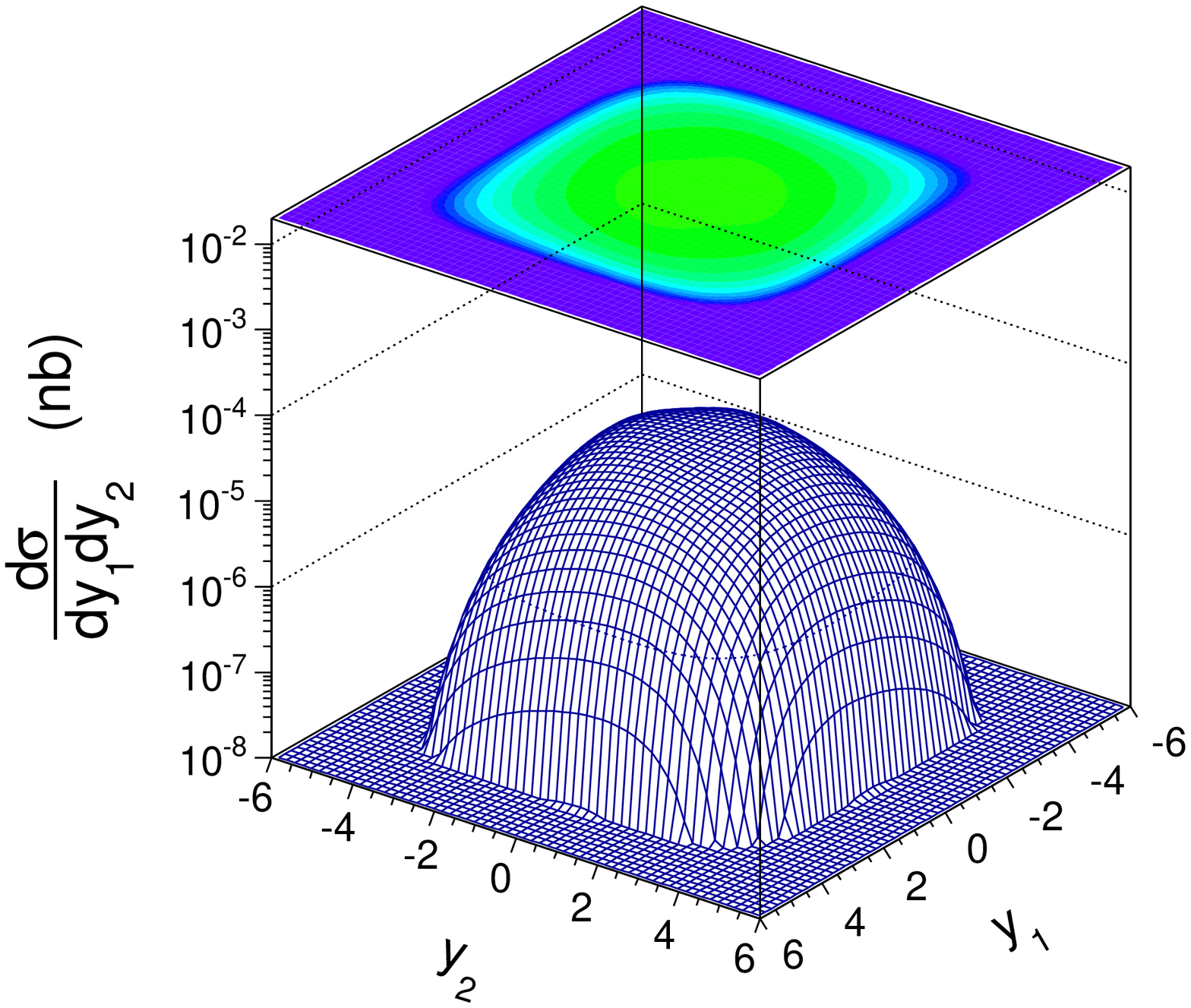}
\includegraphics[width=4.1cm]{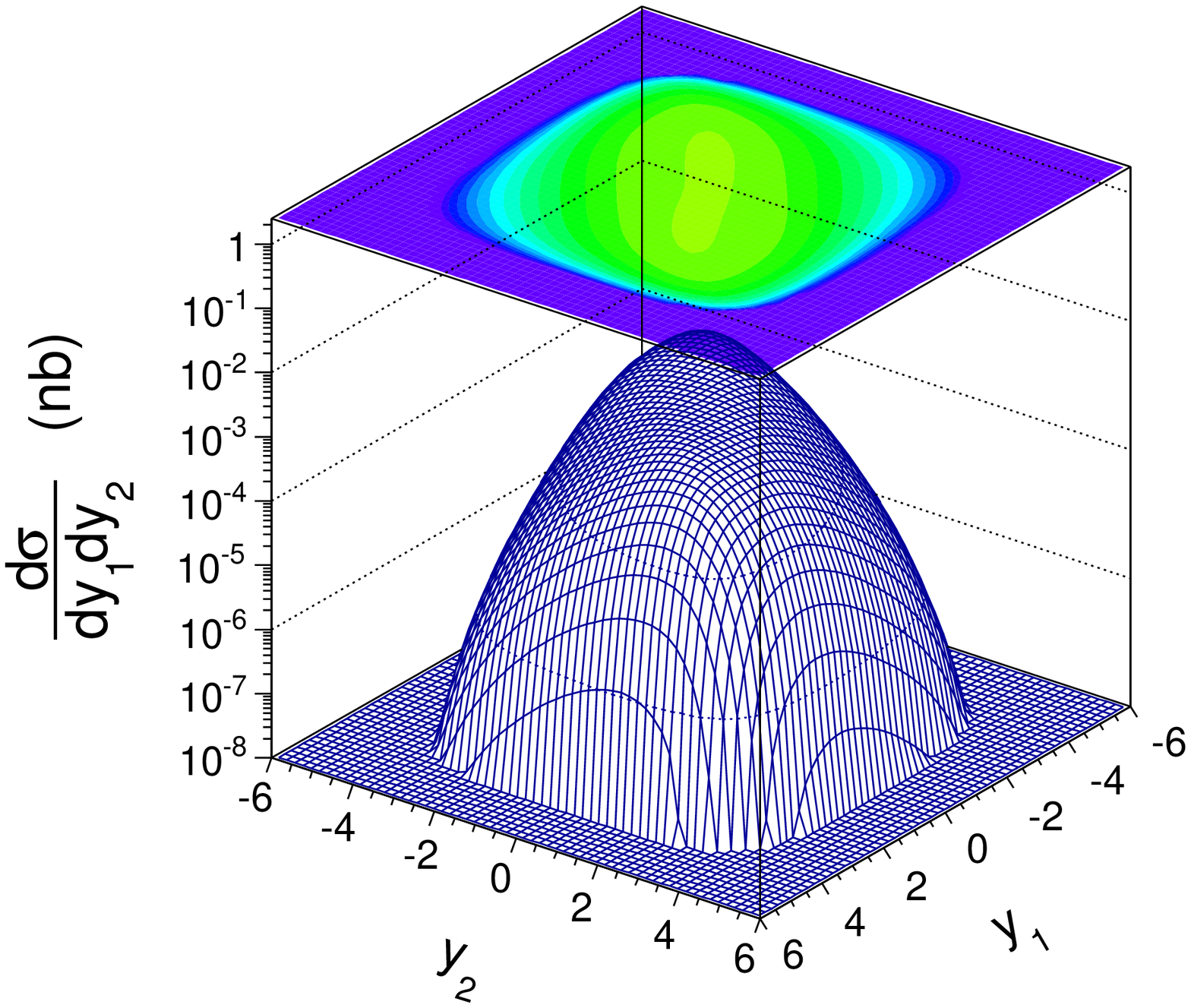}
\end{center}
\caption{
Two-dimensional distributions in rapidity of $W^+$ and rapidity of $W^-$
for the DPS mechanism (left), $\gamma \gamma$ (middle) and $q \bar q$
(right) mechanism for $\sqrt{s}$ = 8 TeV.
}
\label{fig:WW_maps}
\end{figure}

It was argued that the DPS contribution for inclusive $W^+ W^-$ could be
large \cite{KP2013}. Here we partly report results from Ref.~\cite{LSR2015}.
In this analysis we have assumed $\sigma_{eff}$ = 15 mb
as is a phenomenological standard for many other, mostly gluon-gluon
induced, processes.
Similar value was used also in other recent analysis \cite{GL2014}
where in addition evolution effects of dPDFs were discussed.
In our opinion the normalization of the cross section may be an open issue \cite{LSR2015}.
Therefore below we wish to compare rather shapes of a few distributions.
In Fig.~\ref{fig:WW_maps} we show two-dimensional distributions
in rapidity of $W^+$ and $W^-$. For reference we show also
distributions for $\gamma \gamma$ and $q \bar q$ components (see a detailed discussion in Ref.~\cite{LSR2015}).
The DPS contribution seems broader in the $(y_{W^{+}},y_{W^{-}})$ space 
than the other two contributions.

\begin{figure}
\begin{center}
\includegraphics[width=7.0cm]{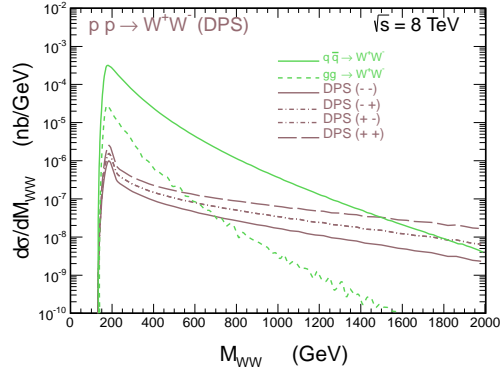}
\end{center}
\caption{$M_{WW}$ invariant mass distribution for different mechanism
discussed in Ref.~\cite{LSR2015}.}
\label{fig:dsig_dM_DPS}
\end{figure}

In Fig.~\ref{fig:dsig_dM_DPS} we show invariant $M_{WW}$ mass 
distribution for $\sqrt{s}$ = 8 TeV. The DPS contribution
seems to dominate at very large invariant masses.

\begin{table}[tb]
\caption{Cross section for $W^+ W^-$ production at different collision energies
for the dominant $q \bar q$ and DPS contributions.}
\label{table1}
\begin{center}
\begin{tabular}{|l|c|c|}
\hline
       &    $q \bar q$  &     DPS          \\
\hline
8000   &    0.032575    &    0.1775(-03)   \\
14000  &    0.06402     &    0.6367(-03)   \\
100000 &    0.53820     &    0.03832       \\
\hline
\end{tabular}
\end{center}
\end{table}

\begin{figure}
\begin{center}
\includegraphics[width=6.0cm]{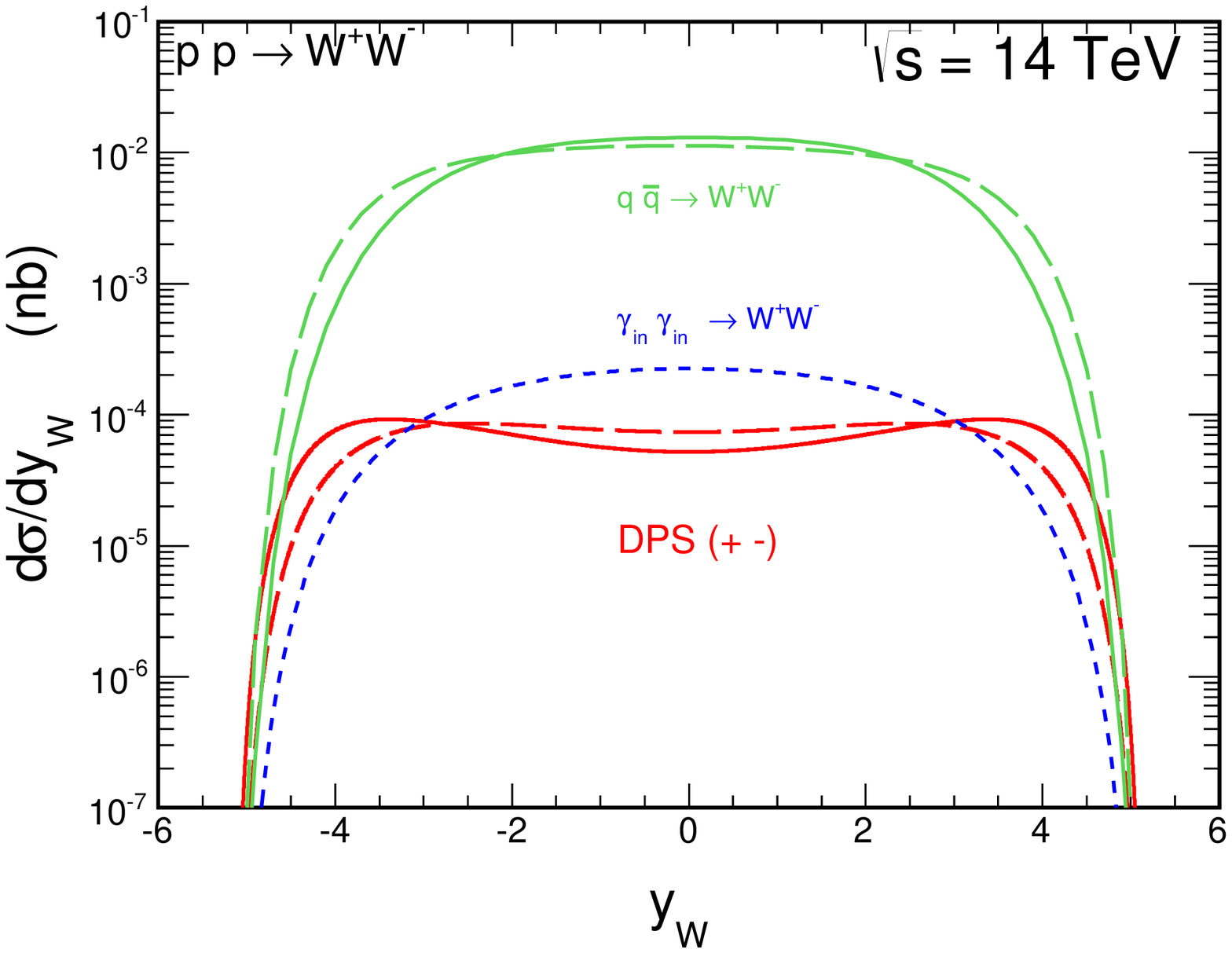}
\includegraphics[width=6.0cm]{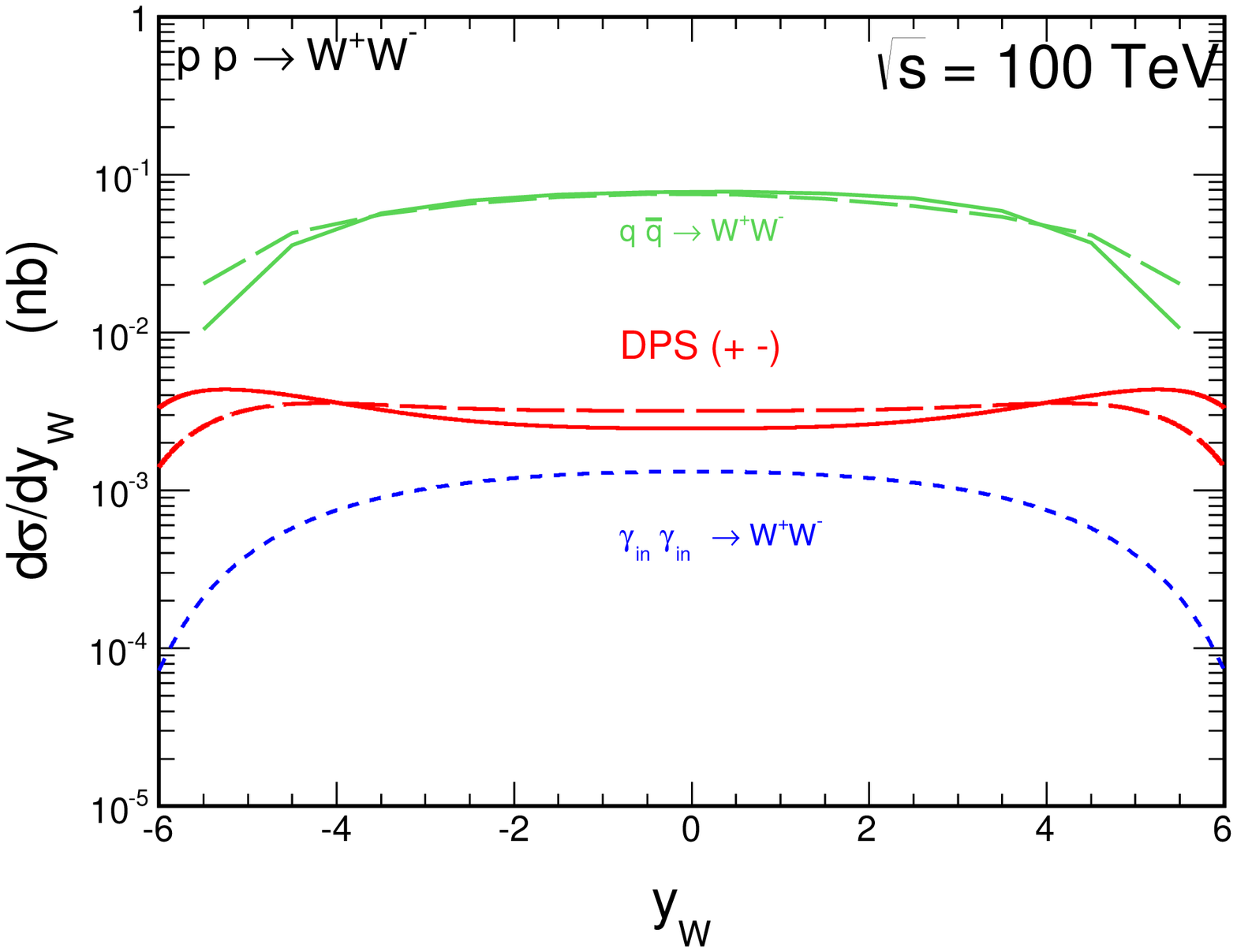}
\end{center}
\caption{Our predictions for future experiments.}
\label{fig:dsig_dy_DPS_future}
\end{figure}

How the situation may look at future high-energy experiments
at the LHC and FCC is shown in Table~\ref{table1} and Fig.~\ref{fig:dsig_dy_DPS_future}. 
Now the DPS (conservative estimation) is relatively larger compared to other contributions.

\begin{figure}
\begin{center}
\includegraphics[width=5.0cm]{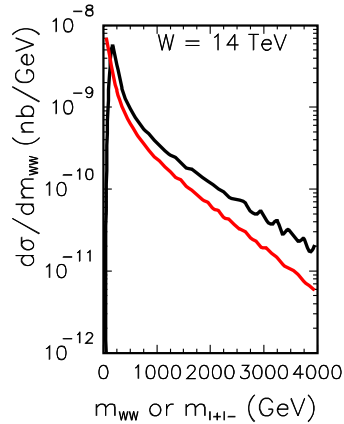}
\end{center}
\caption{Invariant mass distribution of the $W^{+}W^{-}$ system (thick solid line) and corresponding distribution for the $\mu^{+}\mu^{-}$
system. No branching fractions are included.}
\label{fig:dsig_dMWW_lplm}
\end{figure}

In experiments one can measure charged leptons and not $W^{\pm}$
bosons. Therefore a detailed study of lepton distributions is needed.
As en example we show (see Fig.~\ref{fig:dsig_dMWW_lplm}) 
distribution of invariant mass of charged leptons
compared with that for gauge bosons. Only a relatively small shift
towards smaller invariant masses is observed. A more detailed studies 
are necessary to answer whether the $W^+ W^-$ distribution can 
be identified experimentally.
Several background contributions have to be considered.
We leave such a detailed studies for future.

\section{Conclusions}

We have briefly review some double-parton scattering processes
considered by us recently.

First we have shown, within a leading-order collinear-factorization,
that the cross section for $c \bar c c \bar c$
production grows much faster than the cross section for $c \bar c$
making the production of two pairs of $c \bar c$ production
very attractive in the context of exploring the double-parton scattering
processes. 

We have also discussed production of $c \bar c c \bar c$
in the double-parton scattering 
in the factorized Ansatz with each step
calculated in the $k_t$-factorization approach, i.e. including
effectively higher-order QCD corrections.

The cross section for the same process calculated
in the $k_t$-factorization approach turned out
to be larger than its counterpart calculated in the LO
collinear approach.
 
We have calculated also cross sections for the production of 
$D_i D_j$ (both containing $c$ quark)
and $\bar D_i \bar D_j$ (both containing $\bar c$ antiquark) pairs of mesons.
The results of the calculation have been compared to recent results of
the LHCb collaboration.

The total rates of the meson pair production depend on the unintegrated
gluon distributions. The best agreement with the LHCb data has been 
obtained for the Kimber-Martin-Ryskin UGDF. This approach, as discussed
already in the literature, effectively includes higher-order QCD corrections.

As an example we have shown some differential distributions
for $D^0 D^0$ pair production. Rather good agreement has been obtained
for transverse momentum distribution of $D^0$ $(\bar D^0)$ mesons
and $D^0 D^0$ invariant mass distribution. The distribution in azimuthal
angle between both $D^0$'s suggests that some contributions may be still
missing. The single parton scattering contribution, calculated in the
high energy approximation, turned out to be rather small. In the
meantime we checked that $2 \to 4$ ($gg \to c \bar c c \bar c$) $k_t$-factorization approach
leads to similar results as the collinear approach discussed here
\cite{HMS2015}.
 
We have discussed also a new type of mechanism called parton splitting
in the context of the $c \bar c c \bar c$ production.
Our calculation showed that the parton-splitting contribution 
gives sizeable contribution and has to be included when analysing
experimental data.
However, it is too early in the moment for precise predictions of the
corresponding contributions as our results strongly depend on the values
of not well known parameters $\sigma_{eff,2v2}$ and $\sigma_{eff,2v1}$.
Some examples inspired by a simple geometrical model of colliding
partons have been shown.
A better understanding of the two nonperturbative parameters
is a future task.

We have shown that almost all differential distributions 
for the conventional and the parton-splitting contributions 
have essentially the same shape. This makes their 
model-independent separation extremely difficult.
This also shows why the analyses performed so far could describe
different experimental data sets in terms of the conventional 2v2 
contribution alone. The sum of the 2v1 and 2v2 contributions behaves
almost exactly like the 2v2 contribution, albeit with a smaller 
$\sigma_{eff}$ that depends only weakly on energy, scale and
momentum fractions.
With the perturbative 2v1 mechanism included, $\sigma_{eff}$ increases as
$\sqrt{s}$ is increased, and decreases as $Q$ is increased. 

We have discussed also how the double-parton scattering
effects may contribute to large-rapidity-distance dijet correlations.
The presented results were performed in leading-order
approximation only i.e. each step of DPS was calculated in collinear 
pQCD leading-order. Already leading-order calculation provides quite adequate
description of inclusive jet production when confronted with
recent results obtained by the ATLAS and CMS collaborations.
We have identified the dominant partonic pQCD subprocesses relevant for 
the production of jets with large rapidity distance.

We have shown distributions in rapidity distance between
the most-distant jets in rapidity. The results of the dijet SPS
mechanism have been compared to the DPS mechanism. We have performed
calculations relevant for a planned CMS analysis. The contribution of 
the DPS mechanism increases with increasing distance in rapidity between
jets.

We have also shown some recent predictions of the Mueller-Navelet jets
in the LL and NLL BFKL framework from the literature.
For the CMS configuration our DPS contribution is smaller than 
the dijet SPS contribution even at high rapidity distances and
only slightly smaller than that for the NLL BFKL calculation known
from the literature.
The DPS final state topology is clearly different than that for the
dijet SPS (four versus two jets) which may help to disentangle the 
two mechanisms experimentally. 

We have shown that the relative effect of DPS can be increased
by lowering the transverse momenta.  Alternatively one could study 
correlations of semihard pions distant in rapidity. Correlations 
of two neutral pions could be done, at least in principle, with 
the help of so-called zero-degree calorimeters present at each main detectors 
at the LHC.

The DPS effects are interesting not only in the context how they 
contribute to distribution in rapidity distance but per se.
One could make use of correlations in jet transverse momenta,
jet imbalance and azimuthal correlations to enhance the contribution
of DPS. Further detailed Monte Carlo studies are required to settle 
real experimental program of such studies.
The four-jet final states analyses of
distributions in rapidity distance and other kinematical 
observables was performed by us very recently \cite{MS2015}.

Finally we have discussed DPS effects in inclusive production
of $W^+ W^-$ pairs. We have shown that the relative contribution
of DPS grows with collision energy. In experiments one measures rather
electrons or muons than the gauge bosons. Whether experimental
identification of the DPS contribution in this case is possible
requires a detailed Monte Carlo studies.

\vspace{1cm}

{\bf Acknowledgments}

This presentation is based on common work mostly with
Rafa{\l} Maciu{\l}a and partially with 
Jonathan Gaunt, Marta {\L}uszczak and Wolfgang Sch\"afer.
I am very indebted to Rafa{\l} Maciu{\l}a for help in preparing this manuscript.  


\end{document}